\documentclass[12pt]{article}
\usepackage{amsfonts}
\usepackage{amsmath}
\usepackage{latexsym}
\usepackage{epsfig,epstopdf}
\def\beq{\begin{equation}}
\def\eeq{\end{equation}}
\def\bea{\begin{eqnarray}}
\def\eea{\end{eqnarray}}
\begin{document}

%                          Title
\begin{center}
{\Large \bf  Higgs-radion mixing in stabilized brane world models} \\

\vspace{4mm}

%                      author/address
Edward E.~Boos, Viacheslav E.~Bunichev, Maxim A.~Perfilov, Mikhail
N.~Smolyakov,\\ Igor P.~Volobuev \\
\vspace{4mm}
 Skobeltsyn Institute of Nuclear Physics,
Lomonosov Moscow State University
\\ 119991 Moscow, Russia \\

\end{center}

%                        Abstract
\begin{abstract}
We consider a quartic interaction of the Higgs and Goldberger-Wise
fields, which connects the mechanism of the extra dimension size
stabilization with spontaneous symmetry breaking on our brane and
gives rise to a coupling of the Higgs field to the radion and its
KK tower. We estimate a possible influence of this coupling on the
Higgs-radion mixing and study restrictions on model parameters from
the LHC data.
\end{abstract}

\section{Introduction}
In the present paper we consider an extension of the Standard Model (SM) based on
the Randall-Sundrum model with two branes stabilized by a bulk
scalar field \cite{wise,wolfe,Boos:2004uc,Boos:2005dc}, which is
necessary for the model to be phenomenologically acceptable. A
characteristic feature of this extension is the presence of a
massive scalar radion field together with its Kaluza-Klein (KK) tower.
These fields have the same quantum numbers as the neutral Higgs
field. Thus, the radion field and its excitations can mix with the
Higgs field, if they are coupled.

Originally, a Higgs-radion coupling in  the unstabilized
Randall-Sundrum model arising due to a Higgs-curvature term on the
brane was put forward in  \cite{Giudice:2000av}. Then, such a
coupling and the resulting  Higgs-radion mixing in the case of the
stabilized model were discussed in paper \cite{Csaki:2000zn}
without taking into account the KK tower of higher scalar
excitations. The phenomenology of the Higgs-radion mixing
originating from the Higgs-curvature term was also considered in
view of the discovery of the Higgs-like boson at the LHC
\cite{Higgs-discov}; various assumptions about the masses and the
mixings of the scalar states have been analyzed in papers
\cite{Chacko:2012vm}--\cite{Bhattacharya:2014wha}. In particular,
it was shown that the light radion-dominated state with mass below
or above the observed 125-GeV boson is still not completely
excluded by all the electroweak precision constraints and the LHC data.

Here, we discuss a different mechanism of
Higgs-radion mixing immanent in stabilized brane-world models,
where  a Higgs-radion coupling naturally arises due to spontaneous
symmetry breaking on the brane involving the stabilizing scalar
field. This approach takes into account the influence of the  KK
tower of higher scalar excitations on the parameters of the
Higgs-radion mixing, which turns out to be of importance.

 In principle, the most general mechanism of Higgs-radion
mixing can include both possible Higgs-radion couplings. However,
after the spontaneous symmetry breaking on the brane the
Higgs-curvature term also gives rise to a brane-localized
curvature term that affects the mass spectrum and the couplings to
matter of the graviton KK modes \cite{Davoudiasl:2003zt}; this
effect should be taken into account when examining the
four-dimensional effective theory. For this reason, here we restrict ourselves to the new mechanism of Higgs-radion mixing,
which  has the advantage that it modifies only the scalar sector
of the model and leaves intact the masses and the coupling
constants of the graviton KK excitations. This also allows one to
isolate the effects due only to the new mechanism of the mixing.

\section{Higgs-radion interaction}
A stabilized brane-world model in five-dimensional space-time
$E=M_4\times S^{1}/Z_{2}$  with coordinates $\{ x^M\} \equiv
\{x^{\mu},y\}$, $M = 0,1,2,3,4, \, \mu=0,1,2,3 $, with the coordinate
$x^4 \equiv y, \quad -L\leq y \leq L$ parameterizing the fifth
dimension, is defined by the   action
\begin{equation}\label{actionDW}
S = S_g + S_{\phi+SM},
\end{equation}
where $S_g$ and $S_{\phi+SM}$ are given by
\begin{eqnarray}\label{actionsDW}
S_g&=& - 2 M^3\int d^{4}x \int_{-L}^L dy  R\sqrt{g},\\
\label{actionSM} S_{\phi+SM} &=& \int d^{4}x \int_{-L}^L dy
\left(\frac{1}{2}
g^{MN}\partial_M\phi\partial_N\phi - V(\phi)\right)\sqrt{g}\\
\nonumber & -&\int_{y=0} \sqrt{-\tilde g}V_1(\phi) d^{4}x
+\int_{y=L}\sqrt{-\tilde g}(- V_2(\phi) + L_{SM} )  d^{4}x .
\end{eqnarray}
Here the signature of the metric  $g_{MN}$ is chosen to be
$(+,-,-,-,-)$, $M$ is the fundamental five-dimensional energy
scale, $V(\phi)$ is a bulk scalar field potential,
$V_{1,2}(\phi)$ are brane scalar field potentials,
$\tilde{g}=det\tilde g_{\mu\nu}$, and $\tilde g_{\mu\nu}$  denotes
the metric induced on the branes. The space of extra dimension is
the orbifold  $S^{1}/Z_{2}$, which is realized as the circle of
circumference $2L$ with the points $y$ and $-y$ identified.
Correspondingly, the metric $g_{MN}$  and the scalar field $\phi$
satisfy the orbifold symmetry conditions
\begin{eqnarray}
\label{orbifoldsym}
 g_{\mu \nu}(x,- y)=  g_{\mu \nu}(x,  y), \quad
  g_{\mu 4}(x,- y)= - g_{\mu 4}(x,  y), \\ \nonumber
   g_{44}(x,- y)=  g_{44}(x,  y), \quad
   \phi(x,- y)=  \phi(x,  y).
\end{eqnarray}
The branes are located at the fixed points of the orbifold, $y=0$
and $y=L$, and it is assumed that the SM fields  with Lagrangian
$L_{SM}$ live on the brane at $y=L$.

For this form of the action, the vacuum solution for gravity and
the stabilizing scalar field and the vacuum solution for the SM
fields are independent. If we consider  this brane-world model to
be an indivisible theory, it is reasonable to believe that there
should be a common interconnected vacuum solution for all these
fields. To this end, let us modify action (\ref{actionSM}) so that
it contains a quartic interaction of the stabilizing scalar field
and of the Higgs field, which would connect the stabilization of
the extra dimension size and the SM spontaneous symmetry breaking:
\begin{eqnarray}\label{actionsSM1}
S_{\phi+SM} &=& \int d^{4}x \int_{-L}^L dy \left(\frac{1}{2}
g^{MN}\partial_M\phi\partial_N\phi -V (\phi)\right)\sqrt{g} \\
\nonumber & -&\int_{y=0} \sqrt{-\tilde g}V_1(\phi) d^{4}x
+\int_{y=L}\sqrt{-\tilde g}(-V_2(\phi)+ L_{SM-HP} + L_{int}(\phi,
H))    d^{4}x ,
\end{eqnarray}
where the Lagrangian $L_{SM-HP}$  is the SM Lagrangian without the
Higgs potential, which is replaced by the interaction Lagrangian
\begin{equation}
\label{L_int} L_{int}(\phi, H) = -\lambda\left(|H|^2 -\frac{\xi}{M} \phi^2\right)^2,
\end{equation}
 $\xi$ being a positive dimensionless   parameter. A similar
 interaction Lagrangian quadratic in the stabilizing scalar field
 was discussed in \cite{Volobuev:2011zz}.

The background solutions  for the  metric and the scalar field,
which preserve the Poincar\'e invariance in any four-dimensional
subspace $y=const$, look like
\begin{eqnarray}\label{metricDW}
ds^2&=&  e^{-2A(y)}\eta_{\mu\nu}  {dx^\mu  dx^\nu} -  dy^2 \equiv
\gamma_{MN}(y)dx^M dx^N, \\
\phi(x,  y) &=& \phi(y),
\end{eqnarray}
with $\eta_{\mu\nu}$ denoting the flat Minkowski metric, whereas the
background (vacuum) solution for the Higgs field is standard,
\begin{equation}\label{Higgs_vac}
 H_{vac}= \begin{pmatrix}
0\\
\frac{v}{\sqrt{2}}
\end{pmatrix},
\end{equation}
with all the other SM fields  being equal to zero.

If one substitutes this ansatz into the equations corresponding to action
(\ref{actionDW}), one gets  a relation between the vacuum value of
the Higgs field and the value of the field $\phi$ on the brane at
$y= L$,
\begin{equation}\label{relation}
\phi^2(L) = \frac{M v^2}{2\xi}.
\end{equation}

This means that in such a scenario the Higgs field vacuum
expectation value, being proportional to the value of the
stabilizing scalar field on the TeV brane, arises dynamically as a
result of the gravitational bulk stabilization.

The gravitational background solution follows from the well-known
 system of nonlinear differential equations for functions
$A(y),\phi(y)$ \cite{Csaki:2000zn},
\begin{eqnarray}\label{yd}
&\frac{d V}{d\phi}+\frac{dV_1 }{d\phi}\delta(y) +\frac{dV_2
}{d\phi}\delta(y-L)= -4A'\phi'+\phi''&\\ \nonumber &12M^3
(A')^2+\frac{1}{2}\left(V-\frac{1}{2} (\phi')^2\right)=0& \\
\nonumber
&\frac{1}{2}\left(\frac{1}{2}(\phi')^2+V+V_1\delta(y)+V_2\delta(y-L)
\right)=-2M^3\left(-3A''+6(A')^2\right)&,
 \end{eqnarray}
where the first equation is the equation for the scalar field and
the other two follow from the Einstein equations. Here $'=
{d}/{d{y}}$.

Suppose we have a solution $A(y), \phi(y)$ to this system for an
appropriate choice of the parameters of the potentials such that
the interbrane distance is stabilized and  equal to $L$. This means
that the vacuum energy of the scalar field has a minimum for this
value of the interbrane distance.

Now the linearized theory is obtained by representing the metric,
the scalar and the Higgs field in the unitary gauge as
\begin{eqnarray}\label{metricparDW}
g_{MN}(x,y)&=& \gamma_{MN}(y) + \frac{1}{\sqrt{2M^3}} h_{MN}(x,y),
\\ \label{metricparDW1}
\phi(x,y) &=& \phi(y) + \frac{1}{\sqrt{2M^3}} f(x,y),\\
 H(x)&=& \begin{pmatrix}
0\\
\frac{v+ \sigma(x)}{\sqrt{2}}
\end{pmatrix},
\end{eqnarray}
substituting this representation into action (\ref{actionDW}) and
keeping the terms of the second order  in $h_{MN}$, $f$ and
$\sigma$. The Lagrangian of this action is the standard free
Lagrangian of the SM (i.e., the masses of all the SM fields, except
the Higgs field, are expressed in the same way as usual in terms
of the vacuum value of the Higgs field and the coupling constants)
together with the standard second variation Lagrangian of the
stabilized Randall-Sundrum model \cite{Boos:2005dc} supplemented by an
interaction term of fields $f$ and $\sigma$. The part of the
Lagrangian relevant to the Higgs-radion mixing is
\begin{equation}\label{sigma_f_tems}
\left[- \frac{1}{{2M^3}}\left(\frac{1}{2}\frac{d^2V_2}{d \phi^2}
+  \frac{2\lambda v^2  \xi}{M} \right) f^2 + \frac{2\lambda v^2
\sqrt{\xi}}{M^2}\, f\sigma - \frac{1}{2}2\lambda v^2 \sigma^2
\right]\delta(y-L).
\end{equation}

The bulk scalar field $f$ can be expanded in KK modes. In the case
under consideration we can, in the standard way, find the
equations for the mass spectrum and the wave functions of the KK
excitations of the scalar fields just by replacing
$\frac{1}{2}\frac{d^2\lambda_2}{d \phi^2}\rightarrow
\frac{1}{2}\frac{d^2V_2}{d \phi^2} + \frac{2\lambda v^2 \xi}{M}$
in the formulas of paper \cite{Boos:2005dc} (we note that in this
paper the signature of the metric was $(-,+,+,+,+)$, and for this
reason some formulas of that paper my differ in sign from the
corresponding formulas in the present paper). Namely, in this
paper it was shown that it was most convenient to describe the
scalar degrees of freedom of stabilized brane-world models by the
field $\chi=e^{-2A(y)}h_{44}(x,y)$  related to the field $f$ by
the gauge condition
\begin{equation}\label{gauge_f}
f= - 3M^3\frac{e^{2A}}{\phi'}\chi'.
\end{equation}
The equations of motion and the boundary conditions on the branes
for the field $\chi$ define  wave functions $\{\chi_n(y)\}$,
corresponding to mass eigenvalues $\mu_n^2$. Expanding the field
$f$ in these modes, substituting this expansion into the second
variation Lagrangian and integrating over the extra dimension
coordinate $y$, we get a four-dimensional Lagrangian, in which
there is an interaction between the modes and the Higgs field
coming from the term $ \frac{2\lambda v^2 \sqrt{\xi}}{M^2}\,
f\sigma. $ Using  gauge condition (\ref{gauge_f}), the mode
decomposition of $\chi(x,y)$,
$$
\chi(x,y) = \sum_{n = 0}^\infty \phi_n(x)\chi_n(y),
$$
and the boundary condition for the mode wave function $\chi_n(y)$
at $y=L$ \cite{Boos:2005dc} rewritten in terms of eq.
(\ref{sigma_f_tems}),
\begin{equation}\nonumber
 \left(\frac{1}{2}\frac{d^2V_2}{d
\phi^2} + \frac{\phi^{\prime\prime}}{\phi^\prime}  +
\frac{2\lambda v^2  \xi}{M}\right)\chi_n'  -
\mu_n^2e^{2A}\chi_n|_{y=L}=0,
\end{equation}
the field $f$ can be expressed through $\chi_n(y)$ as follows:
$$
\frac{f(x,L)}{3M^3}= - \frac{\chi^\prime
e^{2A}}{\phi^\prime}|_{y=L}= -
\frac{\chi^\prime}{\phi^\prime}|_{y=L}= -
\sum_{n=1}^{\infty}\frac{\mu_n^2}{\left(\frac{1}{2}\frac{d^2V_2}{d
\phi^2} + \frac{\phi^{\prime\prime}}{\phi^\prime}  +
\frac{2\lambda v^2  \xi}{M}\right) \phi^\prime}\chi_n(L)\phi_n(x),
$$
where we have taken into account that $A(L) = 0$ in order to  have
Galilean coordinates on the brane at $y = L$ \cite{Boos:2004uc}.

Thus, the couplings of the modes to the Higgs field can be written
as
$$
\sum_{n=1}^{\infty} \mu_n^2 a_n \phi_n(x)\sigma(x),
$$
where we have  introduced dimensionless quantities
\begin{equation}\label{parameters_an}
- \frac{6\lambda M v^2   \sqrt{\xi}} {\left(\frac{1}{2}\frac{d^2V_2}{d
\phi^2} + \frac{\phi^{\prime\prime}}{\phi^\prime}  +
\frac{2\lambda v^2 \xi}{M}\right)\phi^\prime}  \chi_n(L) = a_n,
\end{equation}
which are supposed to be positive.

These couplings are proportional to the squared masses of the
modes and can be rather large for large $n$. Below we will see
that it is the dimensionless quantities $a_n$ that are of
importance. They are proportional to the values of the wave
functions of the modes on the brane and it is convenient to
express them in terms of the ratios of the latter and $a_1$:
\begin{equation}\label{ratios}
a_n = a_1 \alpha_n, \quad \alpha_n = \frac{\chi_n(L)}{\chi_1(L)}.
\end{equation}
These ratios must go to zero for large $n$ in order for the
cumulative effect of the radion KK tower to be finite. In paper
\cite{Boos:2005dc} the scalar wave functions $\{\chi_n(y)\}$ have
been found explicitly  in a stabilized brane-world model, where
the warp factor is approximately equal to that of the unstabilized
Randall-Sundrum model, and one can check that the quantities $a_n$
are indeed positive and the ratios $\alpha_n$ really  fall off
fairly quickly for large $n$ in this case. It is rather difficult
to find them explicitly in the stabilized brane-world models,
where the warp factor is different from the exponential of a
linear function, as it is in the Randall-Sundrum model.  A study
of such stabilized brane-world models has been carried out in
papers \cite{Boos:2004uc,Boos:2005dc}, and it was found that they
may also solve the hierarchy problem of the gravitational
interaction, giving rise to the masses of KK excitations in the
$\textrm{TeV}$ energy range. However, the corresponding equations
cannot be solved exactly and should be studied numerically for
each set of the fundamental parameters of the model, which is a
very complicated and laborious task.

Here, we will not carry out such calculations for a specific
stabilized brane-world model, but rather give a qualitative
description of the phenomena due to the interaction of the  Higgs
field with the radion and its KK tower, choosing  the  masses and
the coupling constants in a consistent manner and taking into
account the present-day theoretical and experimental restrictions
on their values.

\section{Higgs-radion mixing and the effective Lagrangian}
The part of the extended SM Lagrangian containing the scalar
fields linearly or quadratically looks like
\begin{eqnarray}
L_{part} &=&\frac{1}{2} \partial_\mu \sigma \partial^ \mu \sigma -
\frac{1}{2} 2\lambda v^2 \sigma^2 + \frac{1}{2}
\sum_{n=1}^{\infty}\partial_\mu \phi_n \partial^\mu \phi_n -
\frac{1}{2} \sum_{n=1}^{\infty}\mu_n^2 \phi_n^2 \\ \nonumber &+&
\sum_{n=1}^{\infty} \mu_n^2 a_n \phi_n\sigma - \sum_{f}
\frac{m_f}{v} \bar \psi_f \psi_f \sigma  - \sum_{n=1}^{\infty}b_n
\phi_n (T_\mu^\mu + \Delta T_{\mu}^{\mu})  \\ \nonumber &+&
\frac{2 M^2_W}{v} W_\mu^- W^{\mu +}\sigma  + \frac{M^2_Z}{v}
Z_\mu Z^\mu\sigma + \frac{M^2_W}{v^2}W_\mu^- W^{\mu
+}\sigma^2 + \frac{M^2_Z}{2v^2}Z_\mu Z^\mu\sigma^2,
\end{eqnarray}
where $ T_{\mu}^{\mu}$ is the trace of the SM energy-momentum
tensor and  $\Delta T_{\mu}^{\mu}$ is the conformal anomaly of
massless vector fields explicitly given by
\begin{equation}
\Delta T_{\mu}^{\mu} =
\frac{\beta(g_s)}{2g_s}G^{ab}_{\rho\sigma}G_{ab}^{\rho\sigma} +
\frac{\beta(e)}{2e}F_{\rho\sigma}F^{\rho\sigma}
 \label{Tr1}
\end{equation}
with  $\beta$ being the well-known QCD and QED $\beta$ functions.

The coupling of the scalar modes to the trace of the
energy-momentum tensor comes from the standard interaction
Lagrangian for the metric fluctuations \cite{Boos:2005dc}, which
results in the expression for the parameters $b_n$ in terms of the
scalar mode wave functions:
\begin{equation}\label{bees}
b_n = \frac{1}{2 \sqrt{8M^3}}\chi_n(L).
\end{equation}
In the case of the lowest scalar mode, the radion, this  parameter
is usually denoted as
\begin{equation}\label{Lambda_r}
b_1 =  \frac{1}{\Lambda_r},
\end{equation}
where $\Lambda_r$ is supposed to be of the order of the
fundamental energy scale $M$. Below we will use $\Lambda_r$ as a
natural scale for the interactions of the radion and for those
induced by its KK tower. For this reason, it is also convenient to
express $b_n$ in terms of $\Lambda_r$ and $\alpha_n$ defined in
(\ref{ratios}) as
\begin{equation}
b_n =  \frac{\alpha_n}{\Lambda_r}.
\end{equation}

In what follows, we consider the phenomenology
of the Higgs boson and the radion in the energy range much lower
than the masses of the radion excitations. In this case, we can
pass to a low-energy approximation for this Lagrangian by dropping
the kinetic terms of the radion excitation fields and integrating
them out, which gives the following  effective Lagrangian for the
interactions of the Higgs and radion fields with the SM fields
\begin{eqnarray}
L_{part-eff}&=&\frac{1}{2} \partial_\mu \sigma \partial^ \mu
\sigma - \frac{1}{2} (2\lambda v^2 - d^2)\sigma^2 + \frac{1}{2}
\partial_\mu \phi_1 \partial^\mu \phi_1 -
\frac{1}{2} \mu_1^2 \phi_1^2 + \mu_1^2 a_1 \phi_1\sigma \\
\nonumber &-&  \frac{1}{\Lambda_r} \phi_1 (T_\mu^\mu + \Delta
T_{\mu}^{\mu}) - \frac{c}{\Lambda_r} \sigma (T_\mu^\mu + \Delta
T_{\mu}^{\mu}) - \sum_{f} \frac{m_f}{v} \bar \psi_f \psi_f \sigma
\\ \nonumber &+& \frac{2 M^2_W}{v}W_\mu^- W^{\mu  +}\sigma +
\frac{M^2_Z}{v}Z_\mu Z^\mu\sigma + \frac{M^2_W}{v^2}W_\mu^-
W^{\mu +}\sigma^2 +  \frac{M^2_Z}{2 v^2}Z_\mu Z^\mu\sigma^2,
\end{eqnarray}
where the new parameters are defined in terms of the old ones as
follows:
\begin{equation}
\label{parameter_c}
d^2 = a_1^2\sum_{n=2}^{\infty} {\mu_n^2 \alpha_n^2},\quad {c} =
a_1 \sum_{n=2}^{\infty} \alpha_n^2.
\end{equation}
We have already mentioned that these series should be convergent.
The  positive parameter $d^2$ can be of the order of $v^2$,
because the mass term $2\lambda v^2 - d^2$ should be positive,
and, within the approach under consideration, the coupling
constant of the Higgs boson self-interaction $\lambda$ can be
larger than in the regular SM case. Possible restrictions on the
parameters $c$ will be discussed below.

Next we turn to  the mass matrix of the fields $\sigma$ and
$\phi_1$, which looks like
\begin{equation}\label{mass_matrix}
{\cal M} =\begin{pmatrix}
2\lambda v^2 - d^2& -\frac{1}{2}\mu_1^2 a_1\\
-\frac{1}{2}\mu_1^2 a_1&\mu_1^2\\
\end{pmatrix}.
\end{equation}
This  matrix  can be diagonalized by a rotation matrix
$$
 {\cal R}^T {\cal M R } = diag(m_h^2, m_r^2),\quad
 {\cal R}
=\begin{pmatrix}
\cos \theta& -\sin \theta\\
\sin \theta&\cos \theta\\
\end{pmatrix},
$$
where the rotation angle $\theta$ is defined by the relation
\begin{equation}\label{mixing_angle}
\tan 2\theta = \frac{\mu_1^2 a_1}{\mu_1^2 - 2\lambda v^2 +
d^2}\,\,.
\end{equation}
It is not difficult to see that it is sufficient to take  this
angle only in the interval $-\pi/4 < \theta < \pi/4$, which is one
full period of the function $\tan 2\theta$.

 The observable masses of the mass eigenstates are given by the
expressions
\begin{eqnarray*}
m_h^2 &=& 2\lambda v^2 - d^2 - \frac{\mu_1^2 - 2\lambda v^2 +
d^2}{2} \left( \sqrt{1+ \frac{4a_1^2}{(\mu_1^2 -2\lambda v^2
+ d^2)^2}} - 1\right) \\
m_r^2 &=& \mu_1^2 + \frac{\mu_1^2 - 2\lambda v^2 + d^2}{2} \left(
\sqrt{1+ \frac{4a_1^2}{(\mu_1^2 -2\lambda v^2 + d^2)^2}} -
1\right),
\end{eqnarray*}
which are written in the form that explicitly shows  that the
smaller mass becomes smaller and the larger one becomes larger due
to the mixing. The latter, in particular, means that the
observable masses $m_r^2$ and  $m_h^2$  cannot be exactly equal in
the presence of the mixing and the mixing  angle $\theta$ is
negative for $m_r^2 < m_h^2$ in accordance with the observation in
paper \cite{Csaki:2000zn}.

The original unobservable parameters of the matrix $\cal M$ can be expressed in terms of the observable parameters $m_h^2$, $m_r^2$ and $\theta$. The correspondence between these sets of parameters is one-to-one, if one puts the mixing angle $\theta = 0$ for $m_h^2 = m_r^2$.  Then the following formula for the parameter $a_1$ can be easily obtained:
\begin{equation*}\label{parameter_a1}
  a_1 = \frac{(m_r^2 - m_h^2)\sin 2
 \theta}{m_r^2 \cos^2 \theta + m_h^2 \sin^2 \theta }.
\end{equation*}
This formula, together with the second formula in
(\ref{parameter_c}), gives an expression for the parameter $c$ in
terms of the physical masses and the mixing angle
\begin{equation}\label{restriction}
 c =  \frac{(m_r^2 - m_h^2)\sin 2
 \theta}{m_r^2 \cos^2 \theta + m_h^2 \sin^2 \theta } \left(\sum_{n=2}^{\infty} \alpha_n^2\right)
\end{equation}
which also depends on the sum of the wave function ratios
$\alpha_n^2$. These ratios are, of course, model dependent, and,
although they should fall off with $n$ in order for the sum to be
convergent, one cannot exclude that, in certain models, several
first ratios can be of the order of unity. Thus, one can
conservatively estimate this sum to be also of the order of
unity,\footnote{For example, in the stabilized model discussed in
paper \cite{Boos:2005dc}, the approximate wave functions of the
scalar modes give rather rough estimates for the value of this
sum in the interval $(0.02, 0.2)$ for the radion mass in the
interval $100\, GeV < m_r < 500\, GeV$ and the mass of its first
KK excitation of the order of 1 TeV.} which gives the
estimate for the parameter $c$
\begin{equation}\label{restriction1}
0 < c <  c_{max} = \frac{(m_r^2 - m_h^2)\sin 2
 \theta}{m_r^2 \cos^2 \theta + m_h^2 \sin^2 \theta },
\end{equation}
which will be used in the subsequent calculations.

The physical mass eigenstate fields $h(x), r(x)$ are easily
expressed in terms of matrix ${\cal R}$ and the original fields as
\begin{eqnarray}\label{phys_states}
h(x) &=& \cos \theta \, \sigma(x) + \sin \theta \,\phi_1(x)\\
\nonumber r(x) &=& -  \sin \theta \,\sigma(x) + \cos \theta
\,\phi_1(x).
\end{eqnarray}
The field $h(x)$ is called the Higgs-dominated field, and the
field $r(x)$ is called the radion-dominated field because
$\cos \theta > |\sin \theta|$ in the interval  $-\pi/4 < \theta <
\pi/4$ (we recall that $\theta < 0$ for $m_r^2 < m_h^2$). The
interaction of these fields with the fields of the Standard Model
is given by the following effective Lagrangian:

\begin{eqnarray}\label{h-r-Lagrangian}
L_{h-r} &=& \frac{1}{2} \partial_\mu h \partial^ \mu h -
\frac{1}{2} m^2_h h^2 + \frac{1}{2}
\partial_\mu r \partial^\mu r -
\frac{1}{2} \mu_r^2 r^2 \\ \nonumber &-&  \frac{(c \cos \theta
+ \sin \theta)}{\Lambda_r} h (T_\mu^\mu + \Delta T_{\mu}^{\mu})
+ \frac{(c\sin  \theta -
\cos\theta)}{\Lambda_r} r(T_\mu^\mu + \Delta T_{\mu}^{\mu})\\
\nonumber&-& \sum_{f} \frac{m_f}{v} \bar \psi_f \psi_f (\cos
\theta \, h - \sin \theta \,r)+ \frac{2 M^2_W}{v}W_\mu^-
W^{\mu +}(\cos \theta \,h - \sin \theta \,r)\\
\nonumber&+& \frac{M^2_Z}{v}Z_\mu Z^\mu(\cos \theta \, h -
\sin \theta \, r) + \frac{M^2_W}{v^2}W_\mu^- W^{\mu
+}(\cos \theta \, h - \sin \theta \, r)^2\\
\nonumber&+& \frac{M^2_Z}{2 v^2}Z_\mu Z^\mu(\cos \theta \, h
- \sin \theta \, r)^2.
\end{eqnarray}

The effective four-dimensional  interaction Lagrangian
(\ref{h-r-Lagrangian}) expressed in terms of the physical
Higgs-dominated  $h(x)$ and radion-dominated $r(x)$ fields
contains their interactions with the SM fields and involves only
five parameters in addition to those of the SM\footnote{We do not
assume {\em a priori} with which state, the Higgs-dominated  or the
radion-dominated, the 125-GeV boson observed at the LHC  is
associated.}: the masses of the Higgs-dominated and
radion-dominated fields $m_h$ and $m_r$, the mixing angle
$\theta$, the (inverse) coupling constant of the radion to the
trace of the energy-momentum tensor of the SM fields $\Lambda_r$
and the parameter $c$ that accommodates the contributions of the
integrated-out heavy scalar modes.  We would like to note here
that, according to formulas (\ref{restriction}) and
(\ref{restriction1}), the parameter $c$ is proportional to its
maximal value $c_{max}$ and therefore depends on the masses $m_h$,
 $m_r$ and the mixing angle $\theta$, which is a particular
feature of the model.

It is interesting to examine what occurs to Lagrangian
(\ref{h-r-Lagrangian}) if the fundamental energy scale $M$ goes
to infinity. In the model under consideration, it is impossible
just to take the limit $M \rightarrow \infty$ because there
exists relation (\ref{relation}) for the model parameters, which
includes the vacuum value of the Higgs field. In the stabilized
brane-world model discussed in \cite{Boos:2005dc}, the radion mass
was found to be proportional to $\phi^2(L)/M^3$. Thus, in order to
keep this mass fixed, we have to take the limit $M \rightarrow
\infty$ and $\xi \rightarrow 0$ so that $M^2 \xi =  const$. One
can see that in this case $a_1$, which is proportional to
$M\sqrt{\xi}$, does not vanish, whereas $\Lambda_r$ defined in
(\ref{bees}) and (\ref{Lambda_r}) goes to infinity. As a result,
in Lagrangian (\ref{h-r-Lagrangian}) the terms   with the
energy-momentum tensor that are  proportional to ${1}/{\Lambda_r}$
vanish, but all the other terms remain because the mixing angle
must not be equal to zero. The situation in this case is rather
similar to the one in the  SM extended  by an extra singlet scalar
\cite{Godunov:2015nea,Robens:2015gla}.  However, there are
differences due to  extra interactions between the Higgs field and
the singlet scalar in these models, arising from the scalar field
potentials that  are absent in our model.

We would also like to note here that, if we formally put  the
parameters $a_1$, $c$  and $\theta$ equal to zero, i.e., consider
the case of the zero mixing, Lagrangian (\ref{h-r-Lagrangian})
becomes just the SM Higgs Lagrangian plus the usual Lagrangian of
the radion interaction with the SM fields via the trace of the  SM
energy-momentum tensor. In the case of a nonzero mixing, there
are  additional terms in this Lagrangian that may lead to certain
changes in the collider phenomenology of the Higgs boson and the
radion.  These issues will be discussed in the next section.

\section{Phenomenological constraints from the LHC}
The effective Lagrangian (\ref{h-r-Lagrangian}) describes the
interactions of the Higgs-dominated $h(x)$ and the
radion-dominated $r(x)$ fields with the Standard Model fields and
with each other. A few natural questions arise. What model
parameter regions are allowed by the present-day experimental
data such as LEP and Tevatron searches, Higgs-like 125-GeV boson
discovery and the Higgs signal strength measurements at various
signatures at the LHC, and the limits from searches for the second
Higgs-like boson at the LHC? Is an interpretation of
the observed boson as a radion-dominated state is still possible
for some model parameters?

Detailed answers to such questions require a corresponding
detailed phenomenological analysis,  which will be considered in a
separate study. In this paper we present constraints coming from the discovery of a 125-GeV
Higgs-like boson at the LHC, measurements of the signal strengths
at various production channels and decay modes, and searches for heavier
Higgs bosons.

The Feynman rules  needed for our study can be easily derived from
Lagrangian (\ref{h-r-Lagrangian}) and are given in the table in
the Appendix. We have omitted the terms coming from off-shell
fermions since the corresponding contributions to physical
processes are exactly canceled out  by additional diagrams with
contact four-point vertices coming from the first-order expansions in
gauge couplings of the SM energy-momentum tensor trace as was
proved in \cite{Boos:2014xha}.

The Feynman rules have been implemented as a ``new model'' in a
special version of the CompHEP code \cite{Boos:2004kh,
Boos:2009un, Boos:2014bca} which also includes a routine for
$\chi^2$ analysis of the Higgs signal strength in a way similar to
those used in  papers \cite{Boos:2013mqa, Boos:2014xza}. The next-to-next-to-leading order (NNLO)
corrections are  taken into account in the CompHEP computations
and in the corresponding analysis by multiplying the involved
vertices by correction factors for each model parameter point such
that the partial and the total SM Higgs decay widths and the Higgs
production cross sections in gluon-gluon fusion (GGF) are exactly
equal to those presented by the LHC Higgs Cross Section Working
Group \cite{Dittmaier:2011ti, Heinemeyer:2013tqa}. Because of the
similarity of the Higgs boson and the radion production and decay
amplitudes including loops \cite{Boos:2014xha}, the same correction
factors have been used for the Higgs- and the radion-dominated
states.

We examine two possible scenarios, where the observed 125-GeV
boson is either the Higgs-dominated state or the radion-dominated
state.

\subsection{Higgs-dominated state at 125 GeV}
\begin{figure}[h]
\begin{minipage}[c]{.50\textwidth}
\includegraphics[width=3.4in]{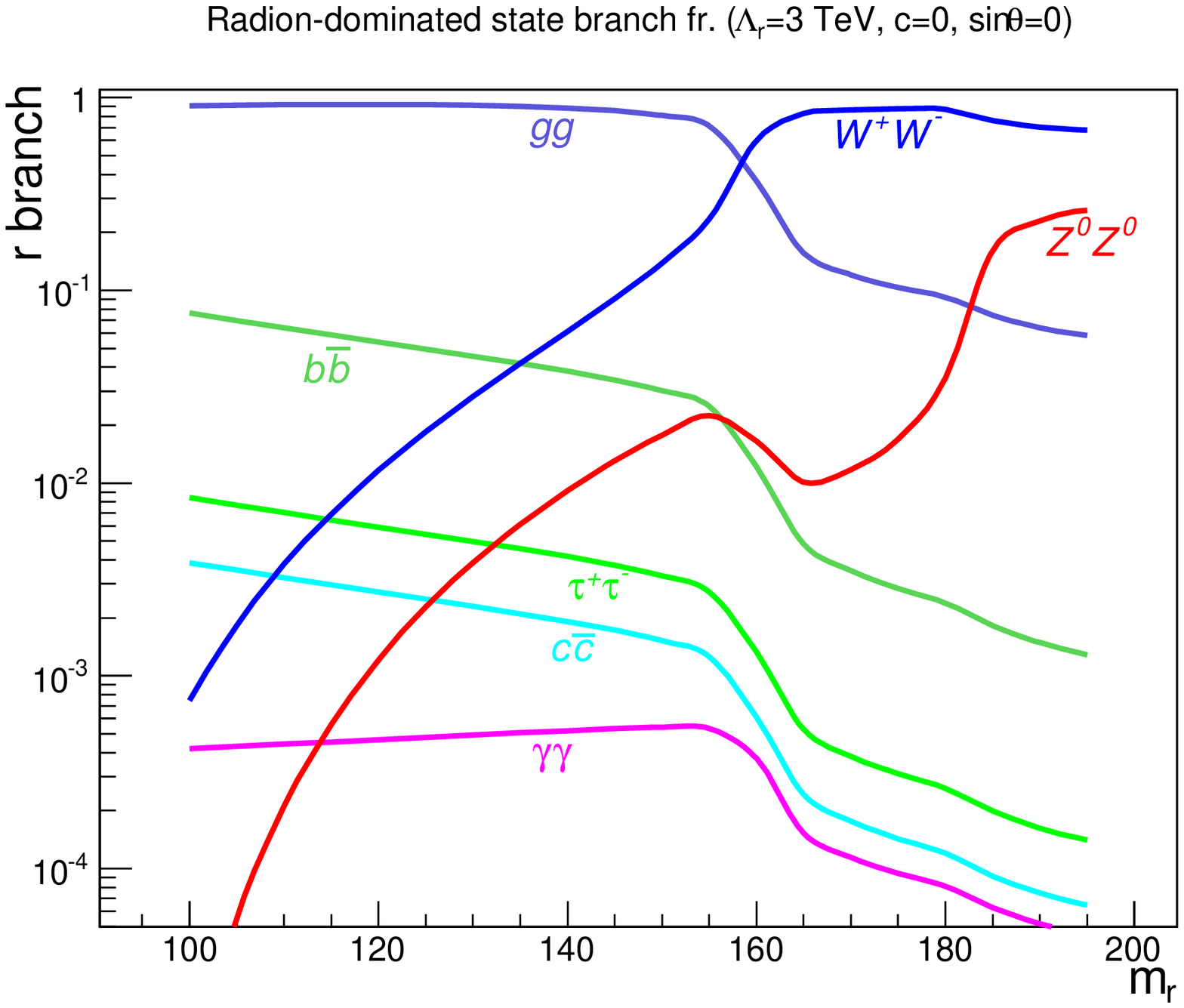}
\end{minipage}
\begin{minipage}[c]{.50\textwidth}
\includegraphics[width=3.4in]{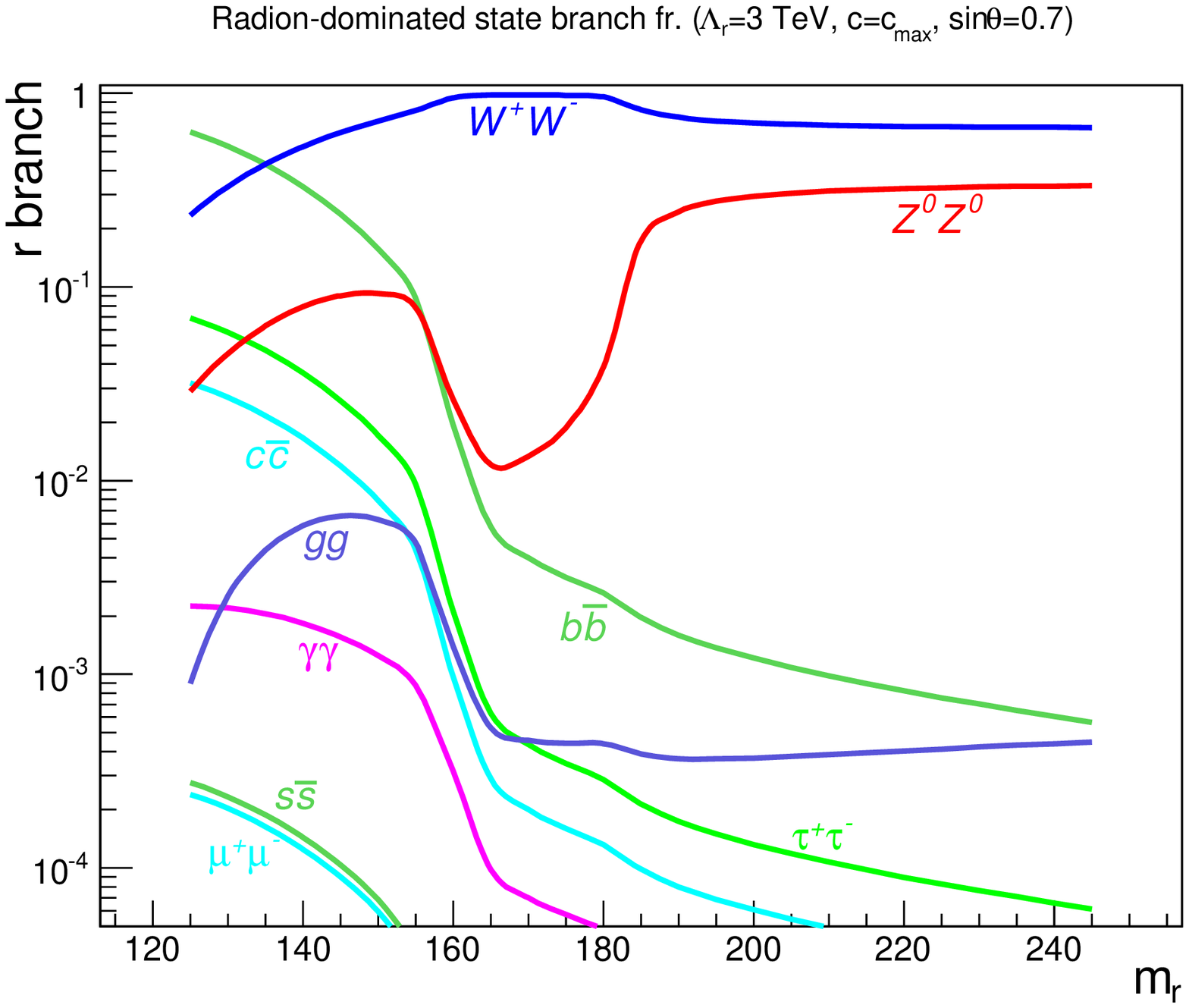}
\end{minipage}
\vspace*{-0.5cm}
 \caption[]{\label{fig_1} Radion-dominated state
branching ratios as functions of $m_r$ ($m_h=125\, GeV$,
$\Lambda_r=3\, TeV$). The left plot corresponds to $\sin\theta=0$
and $c = 0$; the right plot corresponds to $\sin\theta=0.7$, and
$c$ is equal to its maximum value $c_{max}$.}
\end{figure}

In order to understand how much room is left for the
radion-dominated state, we first demonstrate various decay and
production properties of this state. In Fig.~\ref{fig_1}, the decay
branching ratios for the radion-dominated boson are shown  as
functions of its mass.
%*************************************************
%                                Fig.1

The left plot shows the well-known branching behavior
\cite{Cheung:2000rw} of the radion without mixing with the Higgs
boson and interacting with the SM fields only via the trace of the
energy-momentum tensor. As is well known, the main decay mode of
the light radion is the decay to two gluons due to the anomaly
enhancement. However, once the mixing with the Higgs boson is
included, the picture might be changed drastically because of the
appearance of Higgs-like couplings for the radion-dominated state
proportional to $\sin\theta$ and  because of the influence of the
higher scalar modes encoded in the parameter $c$.

\begin{figure}[hH!!!]
\begin{center}
\includegraphics[width=3.4in]{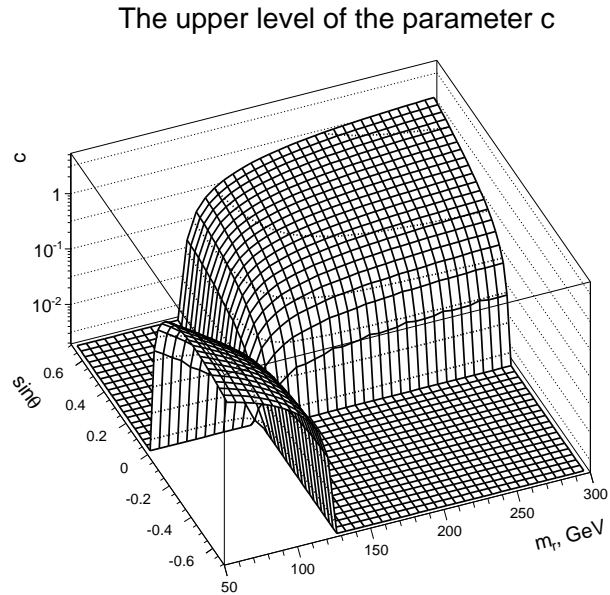}
\end{center}
\vspace*{-0.5cm} \caption[]{\label{fig_2} Two-dimensional plot of
the upper level $c_{max}$ of the parameter $c$ as a function of
$(m_r,\sin\theta)$.}
\end{figure}

\begin{figure}[hH!!!]
\begin{minipage}[c]{.50\textwidth}
\includegraphics[width=3.4in]{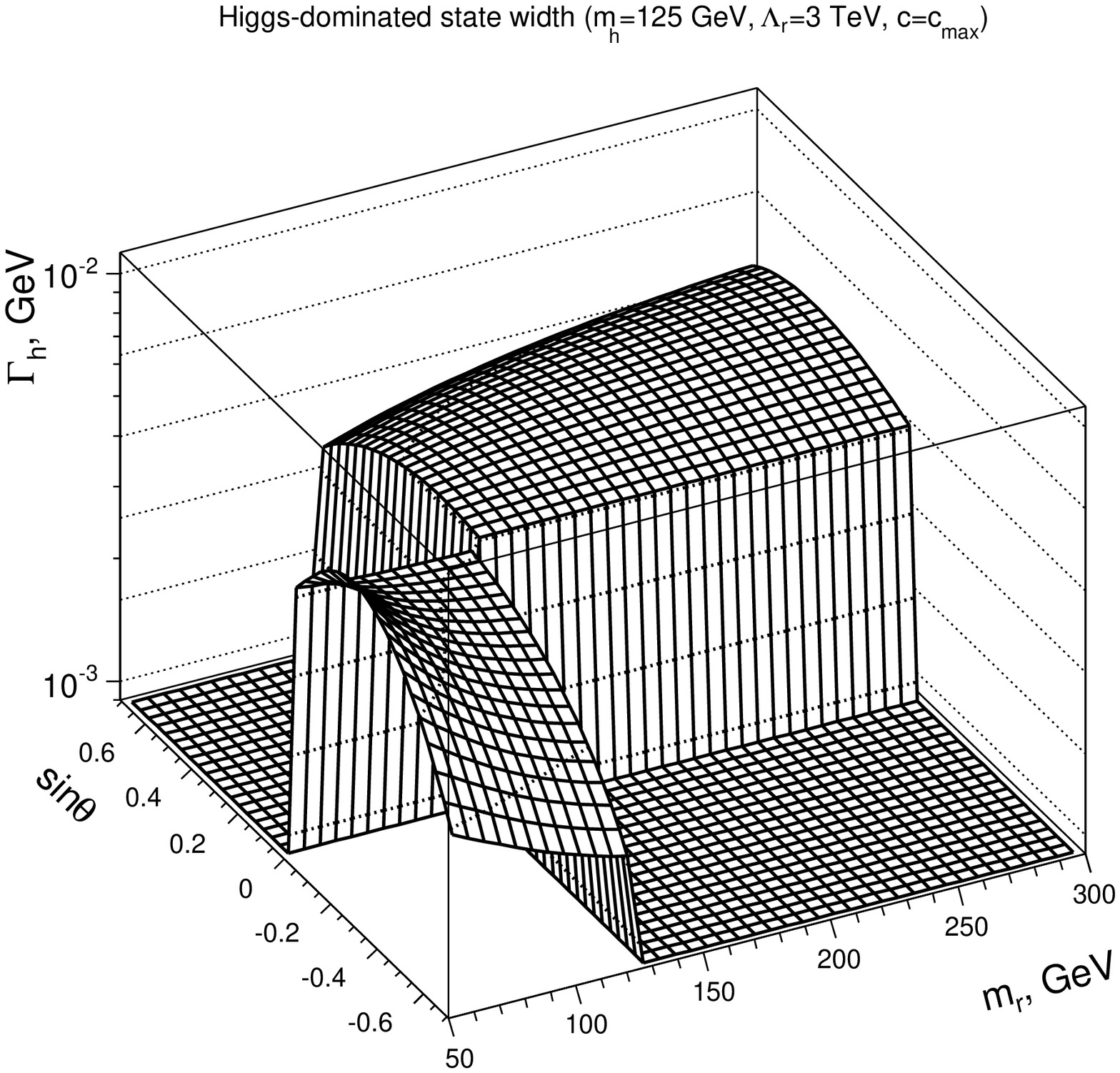}
\end{minipage}
\begin{minipage}[c]{.50\textwidth}
\includegraphics[width=3.4in]{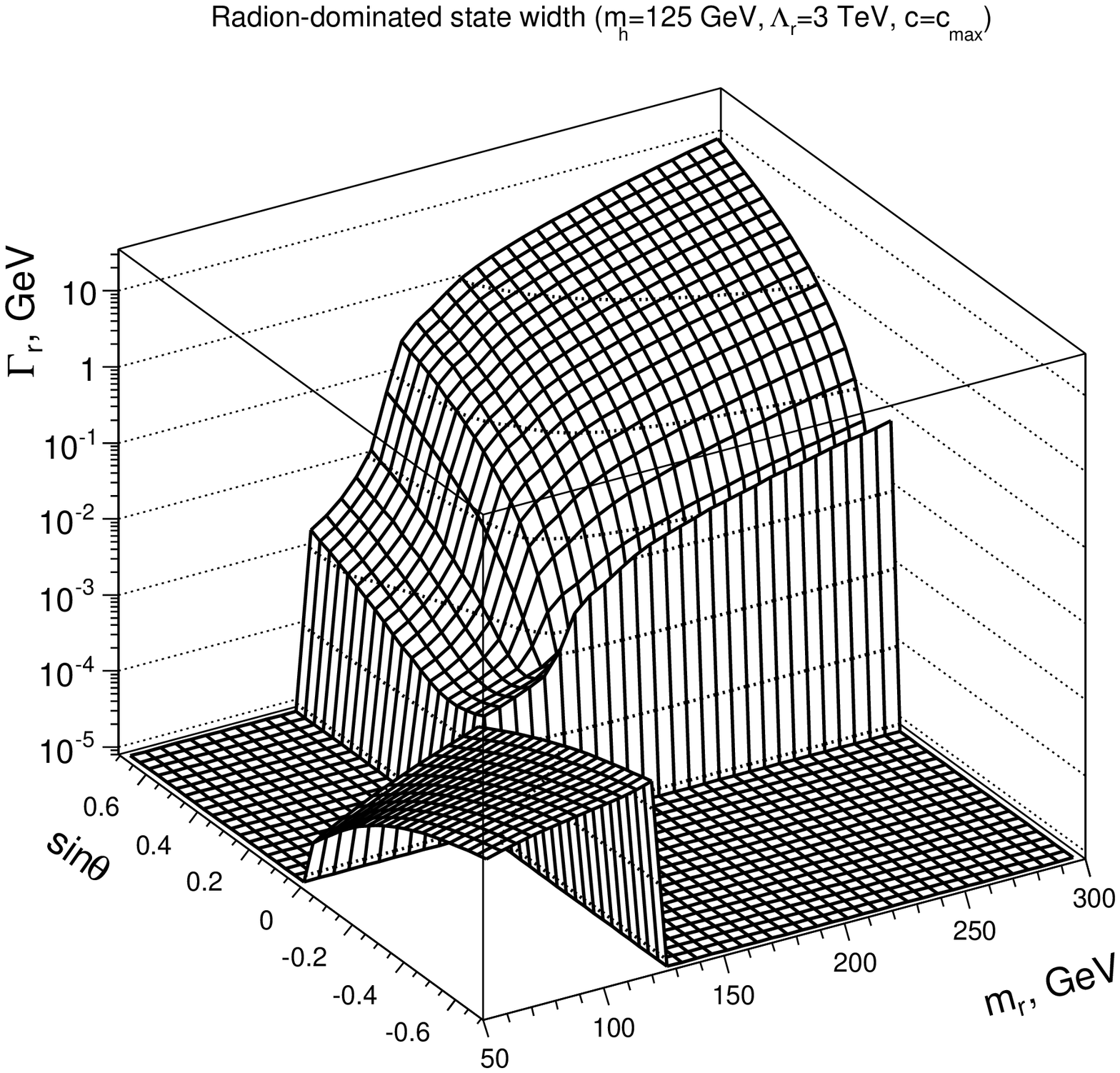}
\end{minipage}
\vspace*{-0.8cm} \caption[]{\label{fig_3} Three-dimensional plots
of the total width as a function of $(m_r,\sin\theta)$ for the LHC
at $\sqrt{s}=8\, TeV$ and  $m_h=125\, GeV$, $\Lambda_r=3\, TeV$,
$c=c_{max}$. The left plot corresponds to the Higgs-dominated
state $h(x)$, and the right plot corresponds to the
radion-dominated state $r(x)$.}
\end{figure}

\begin{figure}[hH!!!]
\begin{minipage}[c]{.50\textwidth}
\includegraphics[width=3.4in]{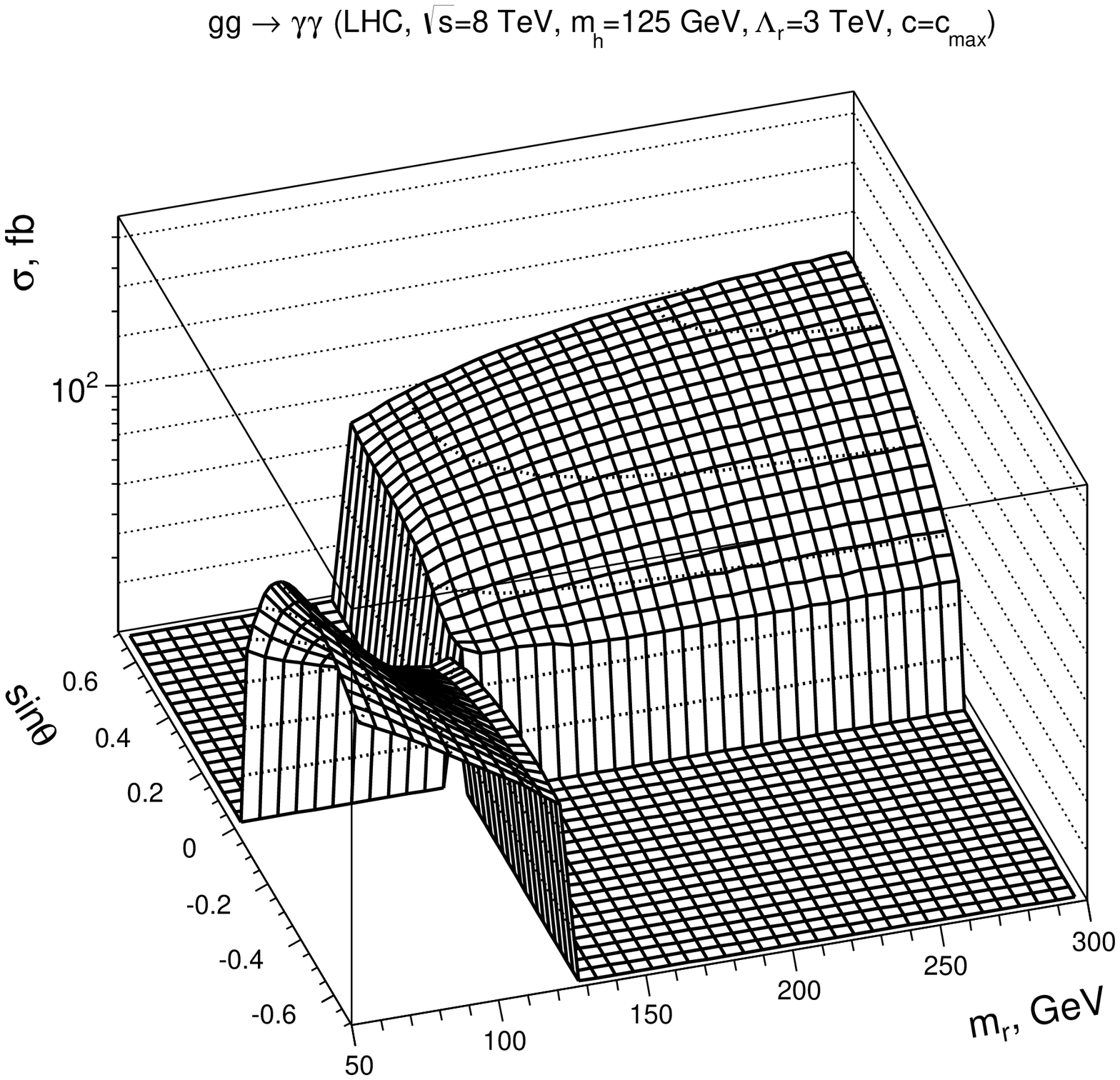}
\end{minipage}
\begin{minipage}[c]{.50\textwidth}
\includegraphics[width=3.4in]{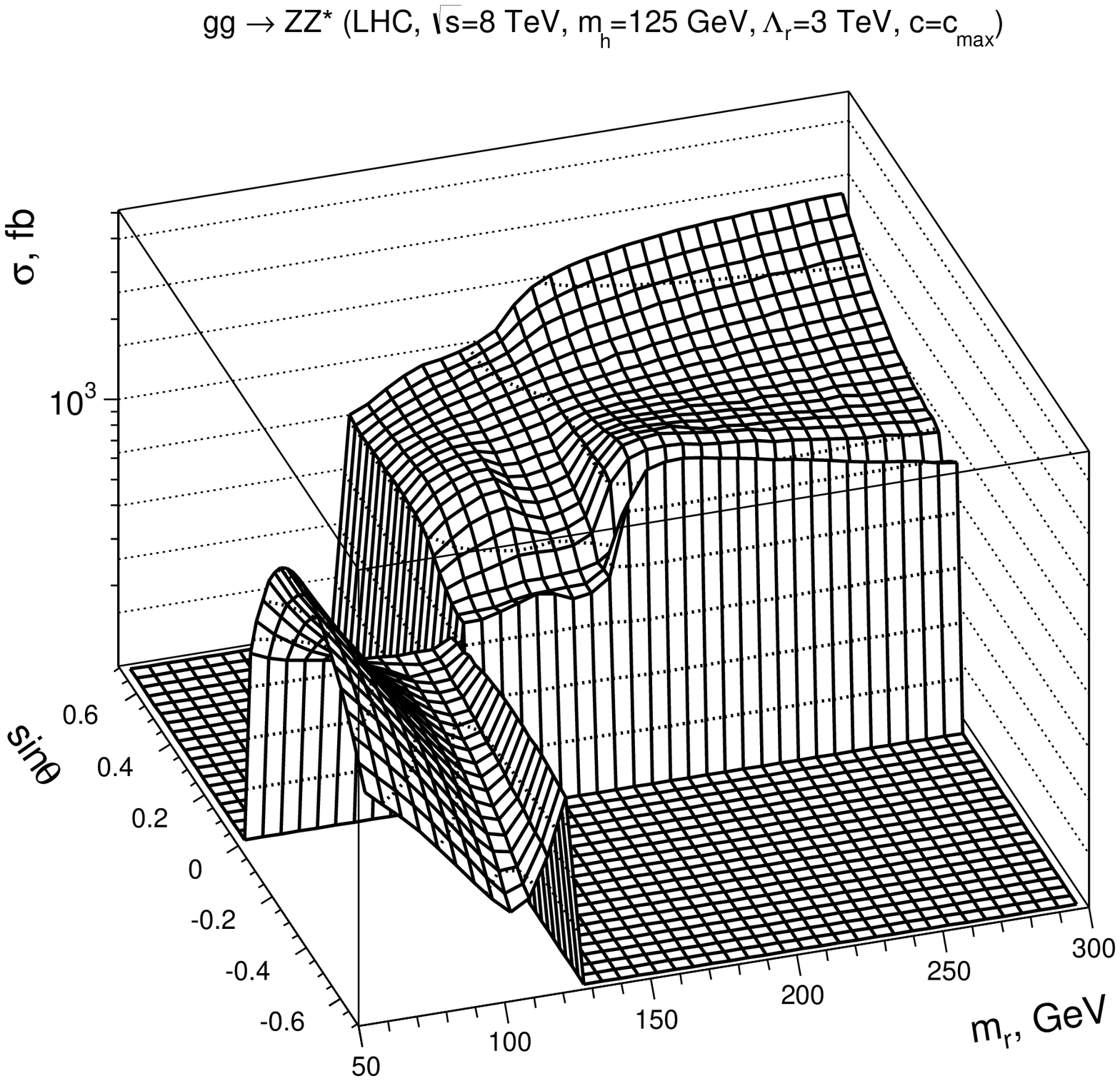}
\end{minipage}
\vspace*{-0.7cm}
\caption[]{\label{fig_4} Three-dimensional plots
of the partial cross section as a function of $(m_r,\sin\theta)$
for the LHC at $\sqrt{s}=8\, TeV$ and  $m_h=125\, GeV$, $\Lambda_r=3\, TeV$,
$c=c_{max}$. The left plot corresponds to the
$gg\rightarrow\gamma\gamma$ channel, and the right plot corresponds
to the $gg\rightarrow ZZ^*$ channel.}
\end{figure}

\begin{figure}[hH!!!]
\begin{minipage}[c]{.50\textwidth}
\includegraphics[width=3.4in]{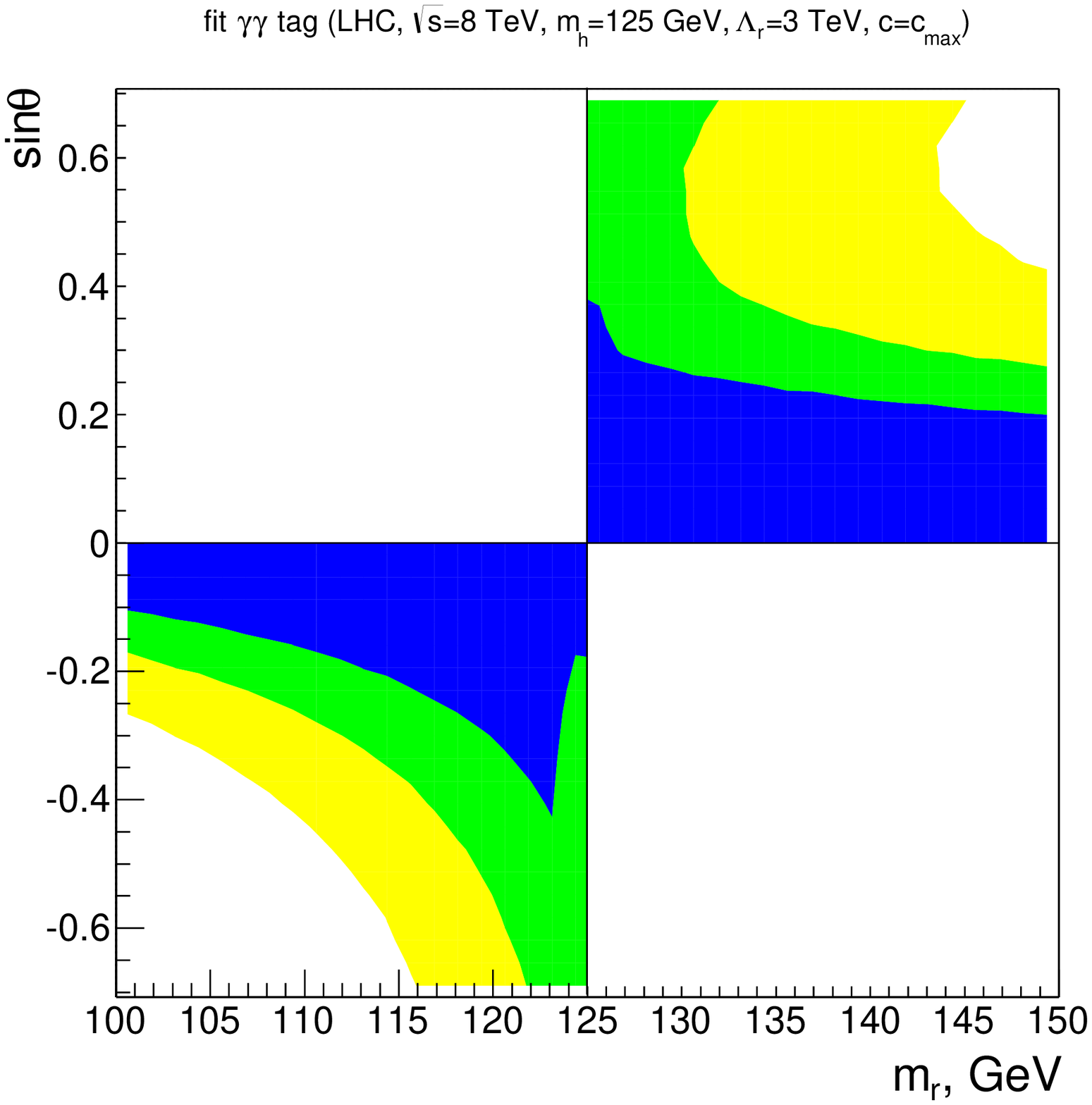}
\end{minipage}
\begin{minipage}[c]{.50\textwidth}
\includegraphics[width=3.4in]{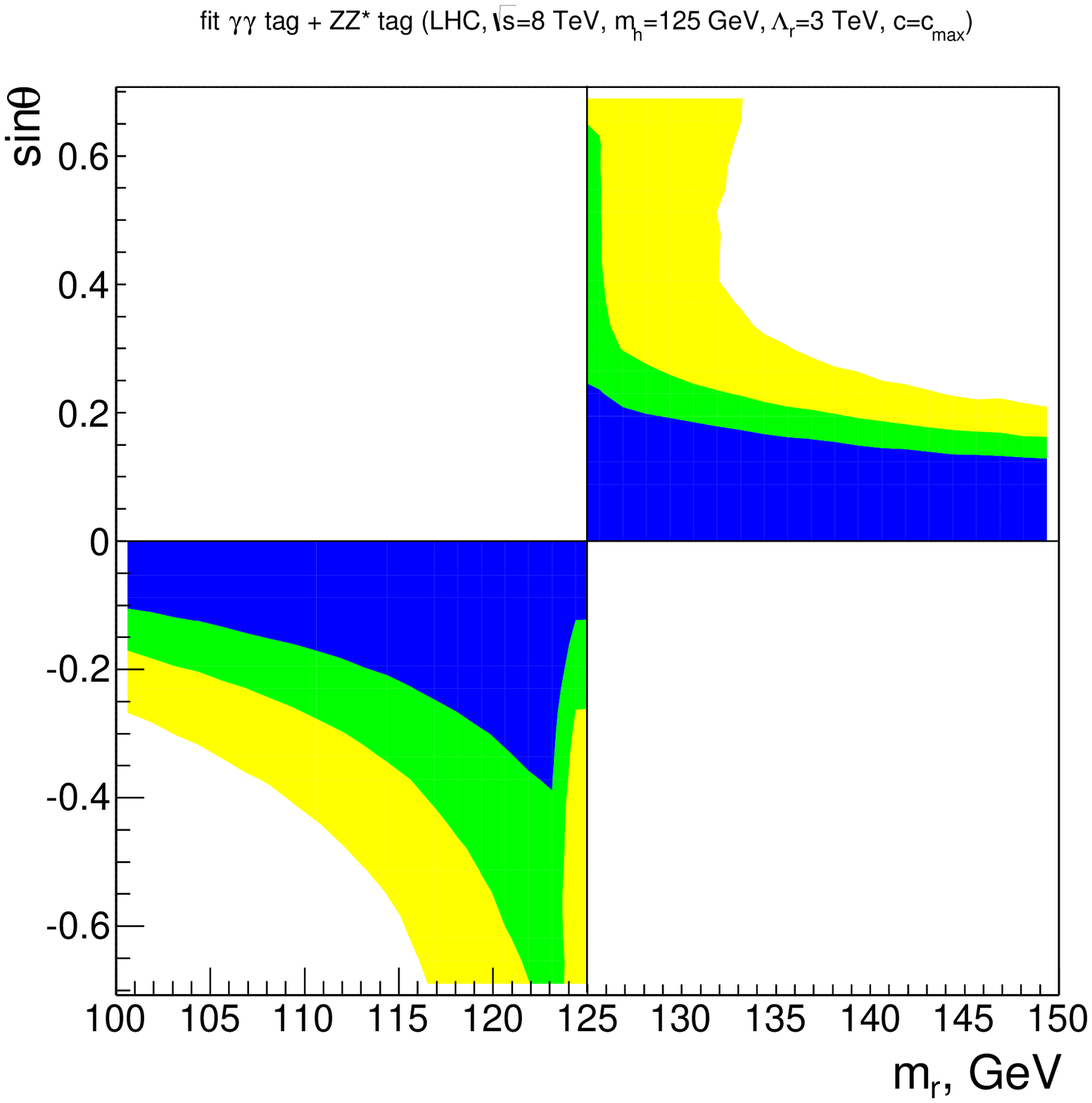}
\end{minipage}
\vspace*{-0.5cm} \caption[]{\label{fig_5} \footnotesize Exclusion
contours for the partial $\chi^2$ fit in the $(m_r,\sin\theta)$
plane for the LHC at $\sqrt{s}=8\, TeV$ and  $m_h=125\, GeV$,
$\Lambda_r=3\, TeV$, $c=c_{max}$. The dark, medium and light shaded
areas correspond to 65\%, 90\% and 99\% CL of the fit,
respectively. The left plot corresponds to the
$pp\rightarrow\gamma\gamma$ channel, and the right plot corresponds
to the combined fit of the $pp\rightarrow\gamma\gamma$ and
$pp\rightarrow ZZ^*$ channels.}
\end{figure}

In Fig.~\ref{fig_2}, the behavior of the upper level of this
parameter  as a function of the radion mass and the mixing angle
is illustrated. We recall that for $\theta > 0$ the
radion-dominated state is heavier than the Higgs-dominated state,
$m_r > m_h$, and vice versa ($\theta < 0$ for $m_r < m_h$), and
therefore, for $\theta < 0, \, m_r > m_h$ and for $\theta > 0, \,
m_r < m_h$, the areas shown as zero plates are not allowed. If one
takes a rather large mixing angle, for example, $\sin\theta=0.7$,
the $b\bar b$ decay mode is dominating as shown in Fig.~\ref{fig_1}
on the right plot. Because of a compensation in the gluon-gluon-radion
vertex between the trace anomaly part proportional to
$1/\Lambda_r$ and the mentioned Higgs-like part proportional to
$\sin\theta /v$ the vertex can be very small for some particular
regions of the parameter space. This leads to an interesting
feature shown in Fig.~\ref{fig_1} that the decay branching ratio
for the radion-dominated state to two gluons may be smaller than
the branching ratio to two photons.
%**************************************************************************
%                                         Fig 2
Because of the mentioned compensations between various parts of
the interaction vertex of the radion-dominated state with gluons,
the behavior of the total width and the production cross sections
for $\gamma\gamma$ and $ZZ^*$ modes in gluon-gluon fusion also
have some minima, as one can see from  Figs.~\ref{fig_3} and
\ref{fig_4}. From the cross sections one gets the corresponding
signal strengths for the radion-dominated state divided by the SM
Higgs cross sections. In our analysis we use the Higgs signal
strengths for all the channels, as given recently by the CMS
\cite{Khachatryan:2014jba} and ATLAS \cite{Aad:2015gba}
collaborations.  As a result of the standard $\chi^2$ analysis,
one gets the regions in the parameter space (the mass of the
radion-dominated state and the mixing angle with the Higgs boson)
which are still allowed. The regions allowed either by considering
only the $\gamma\gamma$ mode, and the $\gamma\gamma$ together with
the  $ZZ^*$ modes are given in the left and the right plots of
Fig.~\ref{fig_5}, correspondingly, for a not-too-large mass range
for the radion-dominated state. For both cases, all the main
production processes [GGF, vector boson fusion (VBF), and
associated production with vector bosons (VH) and with the top
quarks (ttH)] are taken into account. The interference effects of
the Higgs- and radion-dominated states are also taken into
account, which is especially important in the case of close
resonances. The contours in all figures correspond to 65\%, 90\%
and 99\% best-fit confidence level (CL) regions with $\Delta
\chi^2$ less than 2.10, 4.61 and 9.21, respectively, which is the
standard presentation of two-parameter fits and was used in
earlier papers (see, e.g., \cite{Boos:2013mqa,Espinosa:2012ir}).
Thus, the dark shaded area corresponds to 65\% CL of the fit, the
medium shaded area corresponds to 90\% CL of the fit, and the
light  shaded area corresponds to 99\% CL of the fit.

%
%*************************************************
%                                Fig.3
%
%*************************************************
%                                Fig.4
%\clearpage
%
%*************************************************
%                                Fig.5
%=========================================================================
\begin{figure*}[hH!!!]
\begin{center}
\begin{minipage}[t]{.48\textwidth}
\centering
\includegraphics[width=85mm,height=70mm]{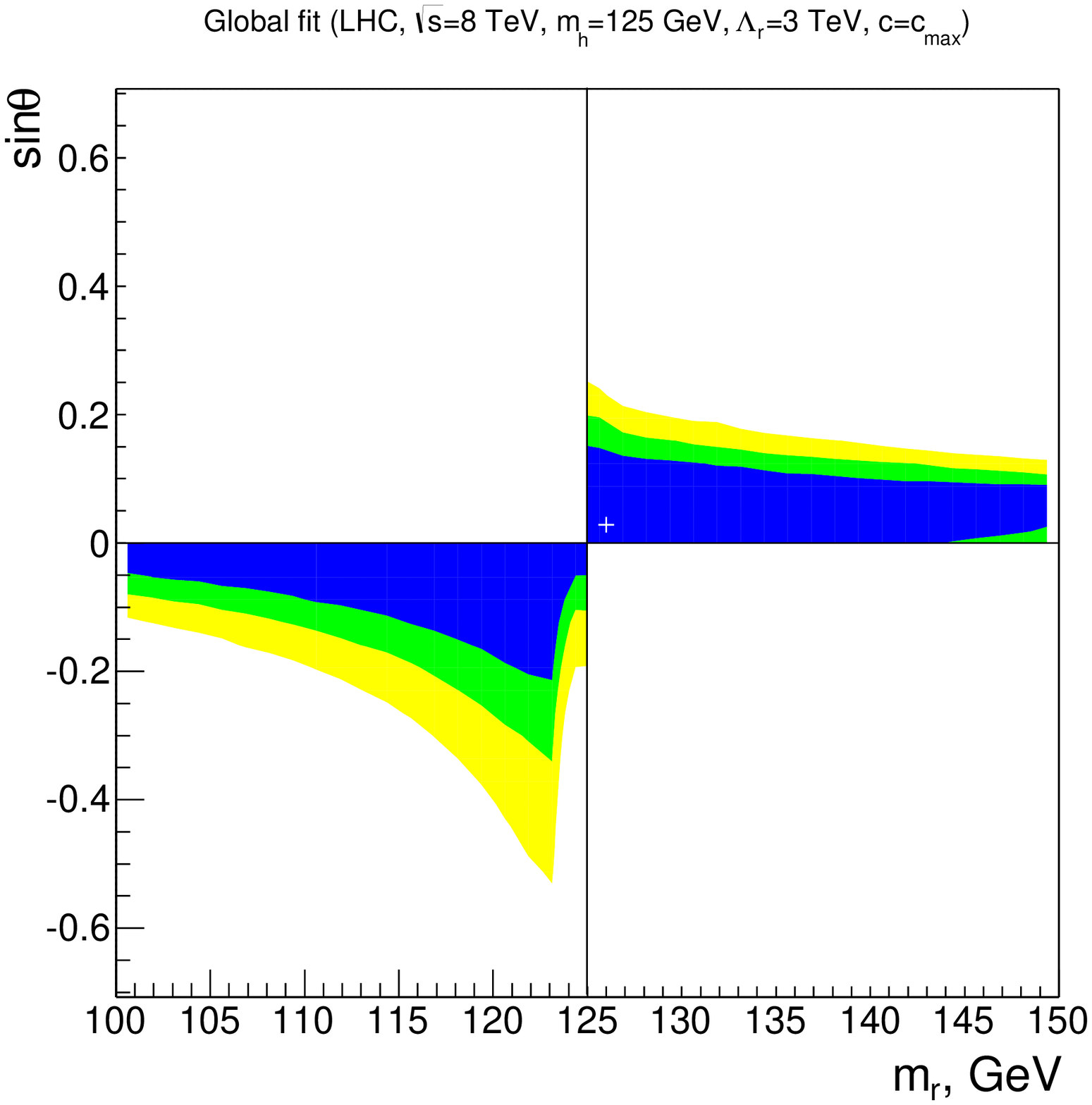}
\vspace*{-1cm} \caption[]{\label{fig_6} \footnotesize Exclusion
contours for the combined $\chi^2$ fit in the $(m_r,\sin\theta)$
plane, which includes all the production processes (GGF, VBF, VH and
ttH) and all the main decay channels ($\gamma\gamma, ZZ^*, WW^*,
b\bar{b}, \tau^+\tau^-$)  for the LHC at $\sqrt{s}=8\, TeV$ and
$m_h=125\, GeV$, $\Lambda_r=3\, TeV$, $c=c_{max}$. The dark, medium and
light shaded areas correspond to 65\%, 90\% and 99\% CL of the fit,
respectively. The white cross marks  the best-fit point.}
\end{minipage}
\hspace{1mm}
\begin{minipage}[t]{.48\textwidth}
\centering
\includegraphics[width=85mm,height=70mm]{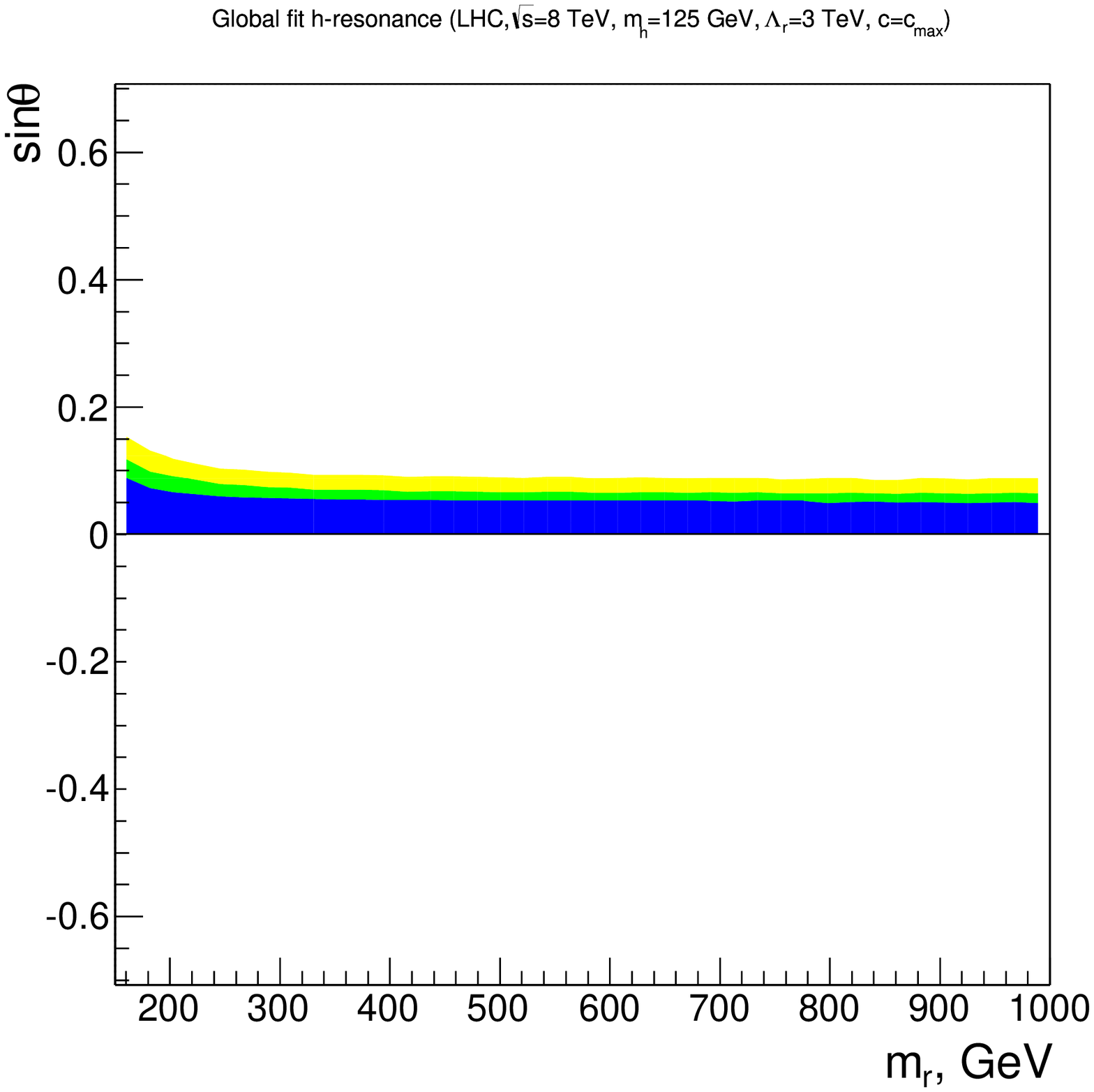}
\vspace*{-1cm} \caption[]{\label{fig_7} \footnotesize Exclusion
contours in the heavy radion mass parameter for the combined
$\chi^2$ fit in the $(m_r,\sin\theta)$ plane that comes from the
signal strengths at 125 GeV for all the production processes (GGF,
VBF, VH and ttH) and all the main decay channels ($\gamma\gamma,
ZZ^*, WW^*, b\bar{b}, \tau^+\tau^-$)  for the LHC at $\sqrt{s}=8\,
TeV$ and $m_h=125\, GeV$, $\Lambda_r=3\, TeV$, $c=c_{max}$. The
dark, medium and light shaded areas correspond to 65\%, 90\% and
99\%  CL of the fit, respectively. }
\end{minipage}
\end{center}
\end{figure*}
%=============================================================================
%
In Fig.~\ref{fig_5}, one can see that the $\gamma\gamma$ mode alone
allows the presence of the radion-dominated state  for masses both
below and above the 125 GeV. The $ZZ^*$ mode gives some
additional restrictions on the parameter space especially in the
mass region closer to the Z-boson pair threshold, where the cross
section is increased. When all the modes are taken into account,
the allowed parameter space region is reduced further, which is
shown in Fig.~\ref{fig_6}.

For larger values of the radion mass, the influence of this
parameter on 125-GeV signal strength is very
 small, resulting in the allowed area shown in Fig.~\ref{fig_7}.
However, in the case of a large resonance mass region, one has to
take into account the exclusion limits given very recently by the CMS
\cite{Khachatryan:2015cwa} and ATLAS \cite{Aad:2015kna}
collaborations, coming from searches for heavy Higgs bosons.
%=========================================================================
\begin{figure*}[!h!]
\begin{center}
\begin{minipage}[t]{.48\textwidth}
%\begin{minipage}[t][2cm][t]{0.4\textwidth}
\centering
\includegraphics[width=85mm,height=70mm]{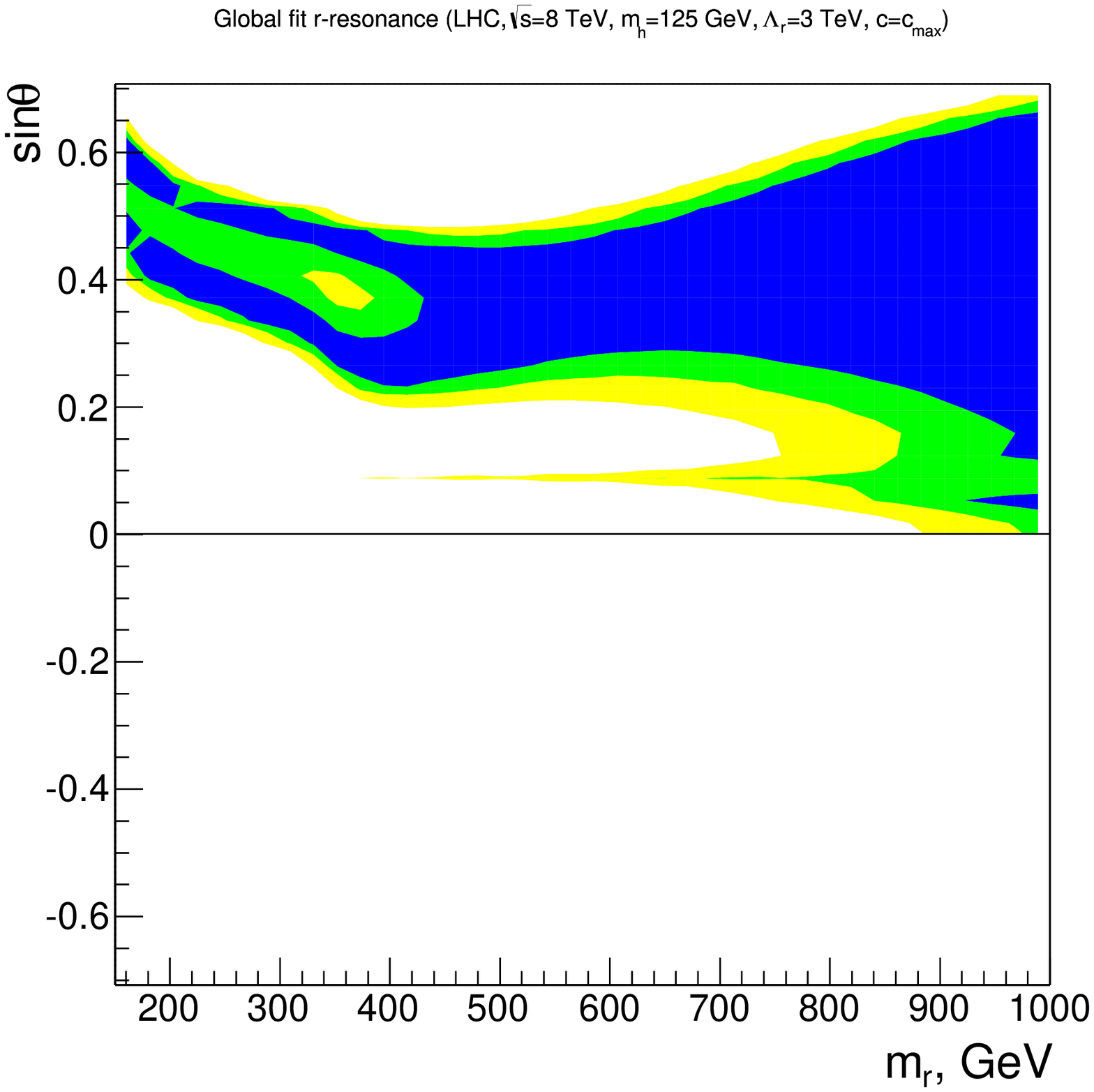}
\vspace*{-1cm} \caption[]{\label{fig_8} \footnotesize Exclusion
contours in the heavy radion mass parameter for the combined $\chi^2$
fit of the CMS and ATLAS exclusion regions at a high mass range by
the resonance production and decay of the radion-dominated state
with corresponding heavy mass in the $(m_r,\sin\theta)$ plane. The
dark, medium and light shaded areas correspond to 65\%, 90\% and 99\% CL of the fit, respectively.}
\end{minipage}
\hspace{1mm}
\begin{minipage}[t]{.48\textwidth}
\centering
\includegraphics[width=85mm,height=70mm]{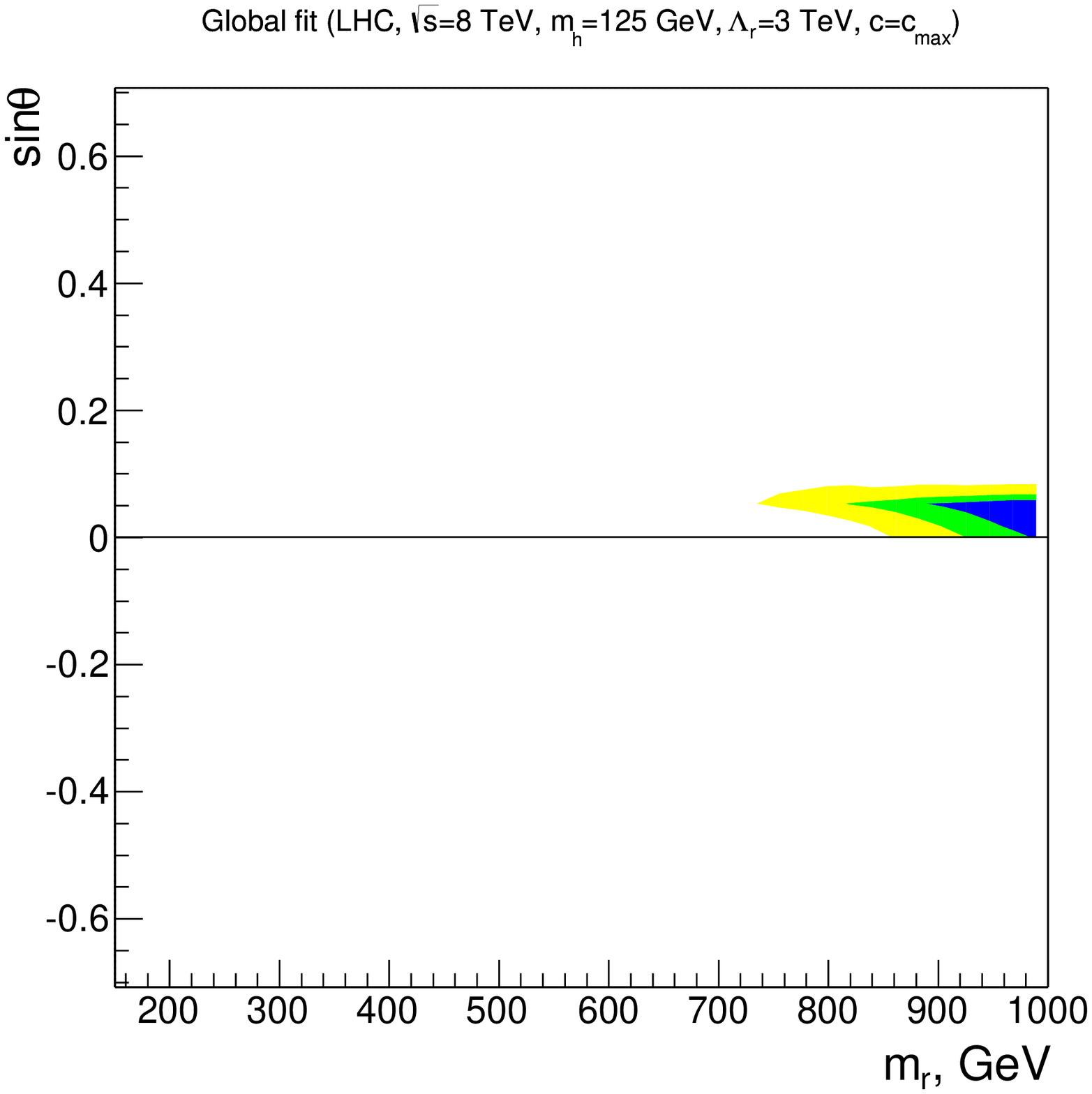}
\vspace*{-1cm} \caption[]{\label{fig_9} \footnotesize The region
in the $(m_r,\sin\theta)$ still allowed by the common fit of the
CMS and ATLAS exclusion limits for heavy Higgs searches and
restrictions from the fit of the influence of those parameters on
the signal strengths at the point 125 GeV. The dark, medium and
light shaded areas correspond to 65\%, 90\% and 99\% CL of the fit,
respectively.}
\end{minipage}
\end{center}
\end{figure*}
%=============================================================================
%
If the radion-dominated state has a large mass, it would
contribute as a resonance in this region. From the comparison of
the radion cross section computation and the experimental limits
by CMS and ATLAS by performing the corresponding $\chi^2$ fit, one
gets a rather large allowed area in the $(m_r,\sin\theta)$
parameter plane, as shown in Fig.~\ref{fig_8}. When taken
together, fits of the data from both the signal strengths and from
the direct searches for heavy Higgs resonances significantly
restrict the allowed region for the heavy radion mass to a small
area at very high masses, as demonstrated in Fig.~\ref{fig_9}.

\begin{figure*}[!h!]
\begin{center}
\begin{minipage}[t]{.48\textwidth}
%\begin{minipage}[t][2cm][t]{0.4\textwidth}
\centering
\includegraphics[width=85mm,height=70mm]{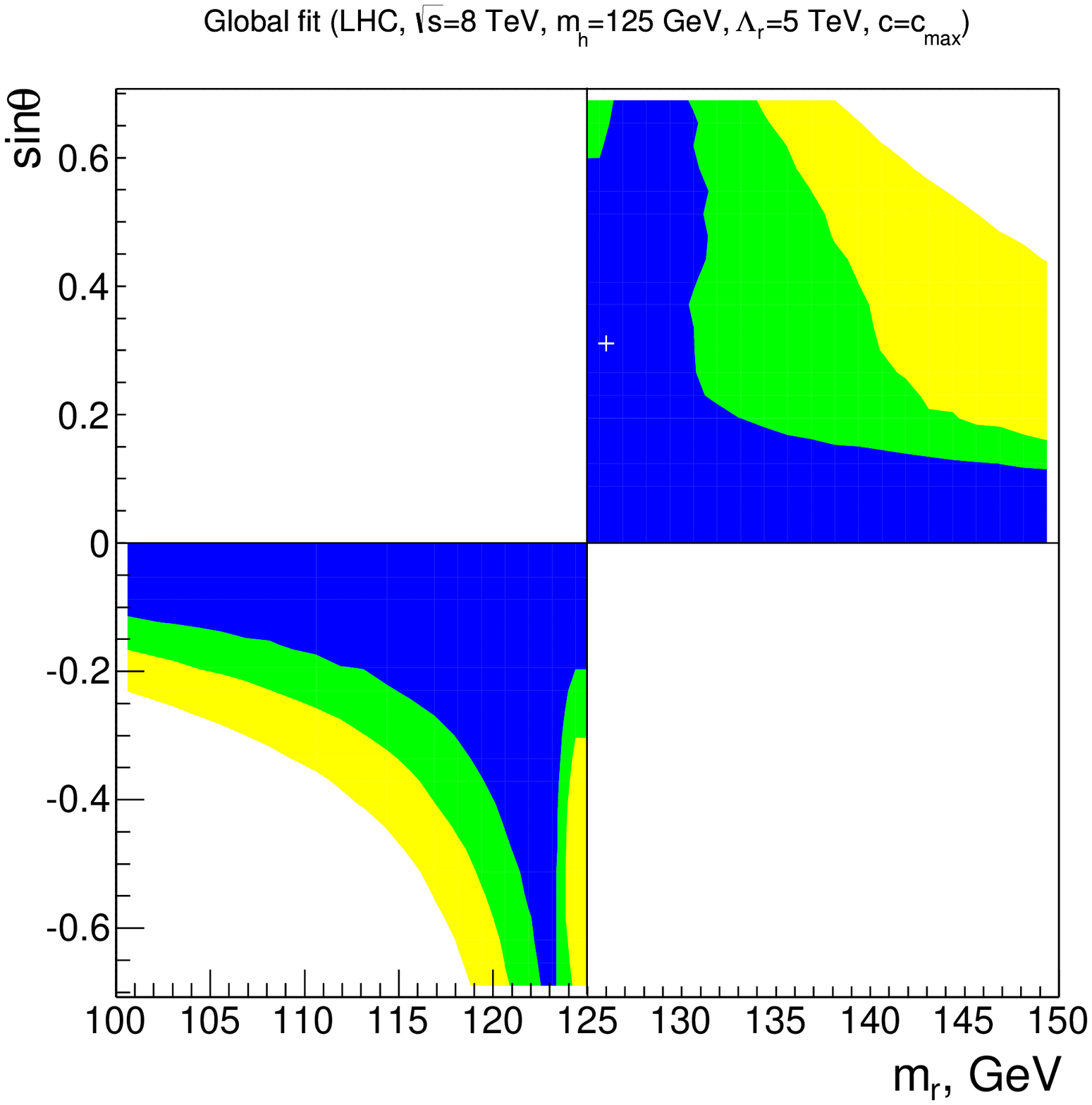}
\vspace*{-1cm} \caption[]{\label{fig_10} \footnotesize Exclusion
contours for the combined $\chi^2$ fit in the $(m_r,\sin\theta)$
plane that includes all the production processes (GGF, VBF, VH and
ttH) and all the main decay channels ($\gamma\gamma, ZZ^*, WW^*,
b\bar{b}, \tau^+\tau^-$)  for the LHC at $\sqrt{s}=8\, TeV$ and
$m_h=125\, GeV$, $\Lambda_r=5\, TeV$, $c=c_{max}$. The dark, medium and
light shaded areas correspond to 65\%, 90\% and 99\% CL of the fit,
respectively. The white cross marks the best-fit point.}
\end{minipage}
\hspace{1mm}
\begin{minipage}[t]{.48\textwidth}
\centering
\includegraphics[width=85mm,height=70mm]{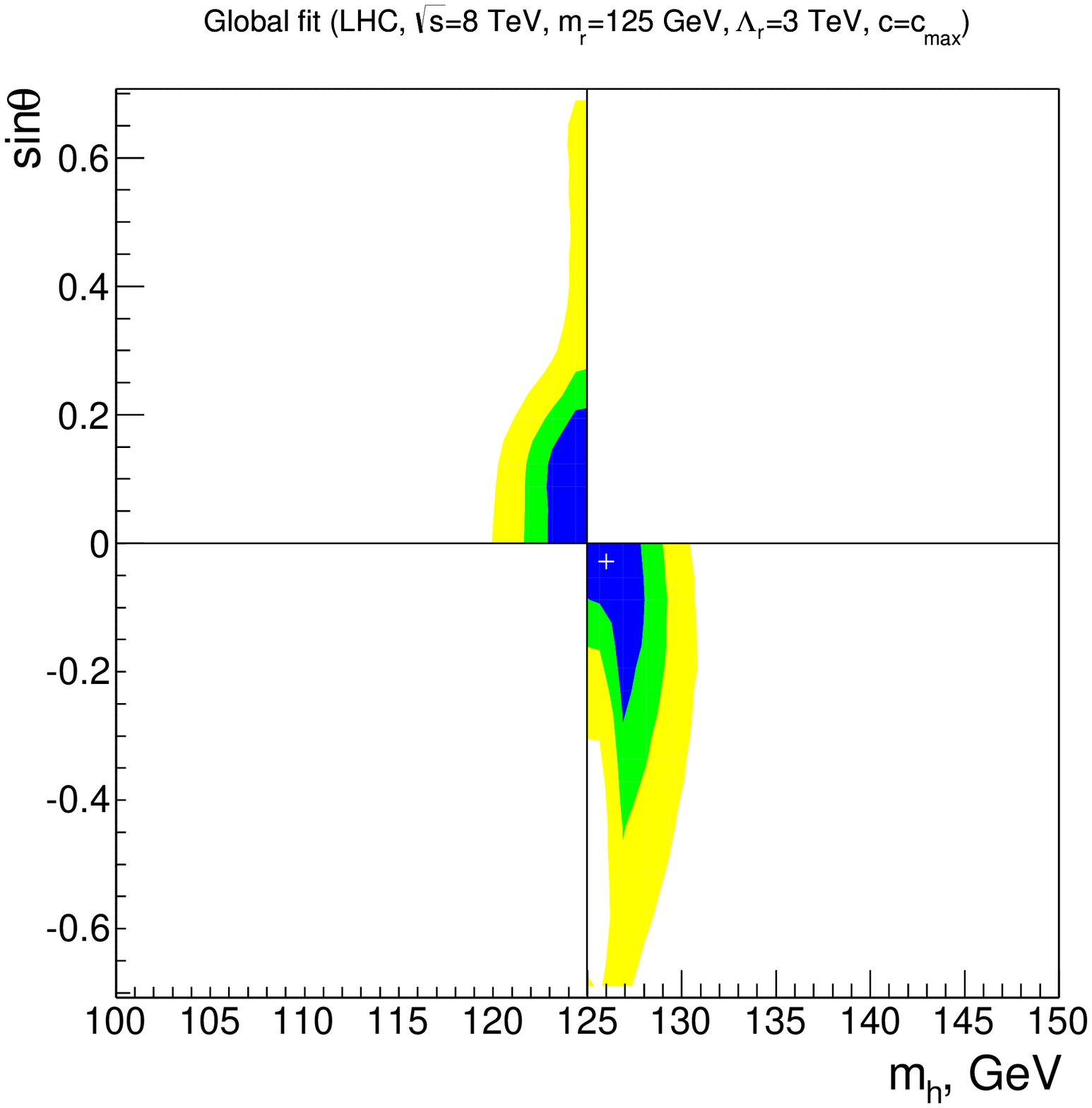}
\vspace*{-1cm} \caption[]{\label{fig_11} \footnotesize
Exclusion contours for the combined $\chi^2$ fit in the $(m_h,\sin\theta)$ plane that
comes from the  signal strengths at 125 GeV for all the
production processes (GGF, VBF, VH and ttH) and all the main decay
channels ($\gamma\gamma, ZZ^*, WW^*, b\bar{b}, \tau^+\tau^-$)  for
the LHC at $\sqrt{s}=8\, TeV$ and $m_r=125\, GeV$, $\Lambda_r=3\, TeV$,
$c=c_{max}$. The dark, medium and light shaded areas correspond 65\%, 90\% and 99\% to
CL of the fit, respectively. The white cross
marks the best-fit point.}
\end{minipage}
\end{center}
\end{figure*}

Obviously, if one considers larger values of the parameter $\Lambda_r$, the cross
section of the radion-dominated state gets smaller and the allowed
region for such a state is increased. This is demonstrated in
Fig.~\ref{fig_10}, where the parameter $\Lambda_r$ is chosen to be
5 GeV with all the other parameters being the same as  for the
considered 3-TeV case.
%=========================================================================
%=============================================================================
%

\subsection{Radion-dominated state at 125 GeV}
Now let us consider briefly the case where the radion-dominated
state has mass 125 GeV, and, using the same analysis strategy,
find the regions for the mass of the Higgs-dominated state and the
mixing angle allowed by two signal strengths. For this case, we
have carried out all the corresponding calculations and have
drawn the plots similar to those in Figs.~\ref{fig_2}--\ref{fig_7}, which we will not present here for the sake of
brevity. We only dwell upon the result of the $\chi^2$ analysis
that is presented in Fig.~\ref{fig_11}.

As one can see, such a scenario is strongly disfavored in
comparison with the previous case. The mass of  the
Higgs-dominated state might still be very close to the mass of the
radion-dominated state  in a wide range of mixing angle. Such a
possibility exists in both scenarios.

\section{Conclusion}

In the present paper we have considered the Higgs-radion mixing
arising in stabilized brane-world models due to merging  the
mechanism of stabilization of the extra dimension size and the
Higgs mechanism of spontaneous symmetry breaking on the TeV brane.
We have discussed phenomenological restrictions on model parameters
coming from coupling measurements and searches for heavy bosons at
the LHC. This mixing is, of course, similar in many aspects to
the one arising due to the Higgs-curvature term on the brane.
However, an important difference is the presence of an extra
coupling at low energies of the Higgs-dominated  field to the
trace of the SM energy-momentum tensor originating from the
coupling of this field to the heavy scalar states of the radion
KK tower.

In order to study the physical consequences of the Higgs-radion mixing in stabilized brane-world models, we derived the effective Lagrangian and gave a qualitative description of the phenomena, taking consistent values for the masses, the coupling constants, and the mixing angle, which are useful for comparing the results of our calculations with the experimental data at the LHC. It turned
out that, though the interaction of an individual higher excited
scalar state with the Higgs field may be weak, their cumulative
effect on the Higgs-radion mixing may be observable. If they give
noticeable contributions to the parameter $c$, it leads to certain
changes in the collider phenomenology of the Higgs boson. A
similar contribution of the directly unobservable higher tensor KK
modes to scattering processes was discussed in \cite{Boos:2007eg}.

In the framework of this model, we have studied two {\em a priori}
possible scenarios: The scalar state discovered at 125 GeV
is either a Higgs-dominated state or a radion-dominated state. In
our analysis, we used measurements of the 125-GeV Higgs signal
strengths by the ATLAS and CMS collaborations and exclusion limits obtained in
searches for heavy Higgs-like states. Our results show that the
interpretation of the 125-GeV scalar state as a
Higgs-dominated state is the preferred  one, although the radion
component in this state can be rather large. Depending on the
value of the radion coupling constant $\Lambda_r$, the allowed
regions for the mass of the radion-dominated state have been
found. It turns out that the radion-dominated state can either
have a mass close to 125 GeV or a mass close to the TeV range.
The allowed regions somewhat increase with the growth of the radion
coupling constant $\Lambda_r$.

We have also shown that the interpretation of the 125-GeV
scalar state as a radion-dominated state is not completely
excluded by the two leading signal strength measurements, though
in this case the restrictions on the allowed masses of the
Higgs-dominated state are very stringent. The mass of the
Higgs-dominated state can be close to 125 GeV, which is in
accordance with our analysis of this state at 125 GeV. Thus, in
the considered model the presence of  two nearly degenerate states
close to 125 GeV is a very probable scenario.

\section{Acknowledgements}
The work was supported by Grant No. 14-12-00363 of the Russian Science
Foundation.

\section{Appendix: Feynman rules}

%=======================================================================
%                               Table 1
\begin{table*}[htb]
\begin{center}
\begin{tabular}{|l|l|} \hline
Triple vertices & Feynman rules \\ \hline $\bar{t}$ \phantom{-}
$t$ \phantom{-} $h$  &
        $-M_t\cdot \left(\frac{C_h}{\Lambda_r}+\frac{\cos\theta}{v} \right)$\\[2mm]
$\bar{t}$ \phantom{-} $t$ \phantom{-} $r$  &
        $-M_t\cdot \left(\frac{C_r}{\Lambda_r}-\frac{\sin\theta}{v} \right)$\\[2mm]
${Z}_{\mu }$ \phantom{-} ${Z}_{\nu }$ \phantom{-} $h$  &
        $2\cdot M_Z^2\cdot \left(\frac{C_h}{\Lambda_r}+\frac{\cos\theta}{v} \right)\cdot g^{\mu\nu}$\\[2mm]
${Z}_{\mu }$ \phantom{-} ${Z}_{\nu }$ \phantom{-} $r$  &
        $2\cdot M_Z^2\cdot \left(\frac{C_r}{\Lambda_r}-\frac{\sin\theta}{v} \right)\cdot g^{\mu\nu}$\\[2mm]
$W^+_{\mu }$ \phantom{-} $W^-_{\nu }$ \phantom{-} $h$  &
        $2\cdot M_W^2\cdot \left(\frac{C_h}{\Lambda_r}+\frac{\cos\theta}{v} \right)\cdot g^{\mu\nu}$\\[2mm]
$W^+_{\mu }$ \phantom{-} $W^-_{\nu }$ \phantom{-} $r$  &
        $2\cdot M_W^2\cdot \left(\frac{C_r}{\Lambda_r}-\frac{\sin\theta}{v} \right)\cdot g^{\mu\nu}$\\[2mm]
${G}_{\mu}$ \phantom{-} ${G}_{\nu}$ \phantom{-} $h$  &
        $\frac{g^2_s}{8\pi^2}\cdot \left[ b_{QCD} \cdot \frac{C_h}{\Lambda_r} + F_{t} \cdot \left(\frac{C_h}{\Lambda_r}+\frac{\cos\theta}{v} \right) \right] \cdot \big(g^{\mu \nu} p_1 p_2
-p_1^\nu p_2^\mu \big)$\\[2mm]
${G}_{\mu}$ \phantom{-} ${G}_{\nu}$ \phantom{-} $r$  &
        $\frac{g^2_s}{8\pi^2}\cdot \left[ b_{QCD} \cdot \frac{C_r}{\Lambda_r} + F_{t} \cdot \left(\frac{C_r}{\Lambda_r}-\frac{\sin\theta}{v} \right) \right] \cdot \big(g^{\mu \nu} p_1 p_2
-p_1^\nu p_2^\mu \big)$\\[2mm]
${A}_{\mu }$ \phantom{-} ${A}_{\nu }$ \phantom{-} $h$  &
        $\frac{e^2}{8\pi^2}\cdot \left[ (b_{2}+b_{Y}) \cdot \frac{C_h}{\Lambda_r} + \left(F_{W}+\frac{8}{3}F_{t}\right) \cdot \left(\frac{C_h}{\Lambda_r}+\frac{\cos\theta}{v} \right) \right] \cdot \big(g^{\mu \nu} p_1 p_2
-p_1^\nu p_2^\mu \big)$\\[2mm]
${A}_{\mu }$ \phantom{-} ${A}_{\nu }$ \phantom{-} $r$  &
        $\frac{e^2}{8\pi^2}\cdot \left[ (b_{2}+b_{Y}) \cdot \frac{C_r}{\Lambda_r} + \left(F_{W}+\frac{8}{3}F_{t}\right) \cdot \left(\frac{C_r}{\Lambda_r}-\frac{\sin\theta}{v} \right) \right] \cdot \big(g^{\mu \nu} p_1 p_2
-p_1^\nu p_2^\mu \big)$\\[2mm]
 \hline
\end{tabular}
\end{center}
\vspace*{-0.5cm} \caption[]{\label{table_1} Triple vertices.}
\end{table*}
%=======================================================================
In Table~1, the constants and the functions of the momenta are
explicitly given by the following expressions: $b_{QCD}=7$,
$b_2=\frac{19}{6}$, $b_Y=-\frac{41}{6}$, $C_h=\sin\theta+c\cdot
\cos\theta$, $C_r=\cos\theta-c\cdot \sin\theta$, $F_W = - (2 +
3y_W + 3y_W (2 - y_W )f(y_W ))$, $F_t = y_t (1 + (1 - y_t )f(y_t
))$, $y_i = 4m^2_i/(2p_1\cdot p_2)$,
\begin{displaymath}
f(z) = \left \{ \begin{array}{cr}
\left[ \sin^{-1} \left(\frac{1}{\sqrt{z}} \right ) \right ]^2\;, & z \ge 1 \\
-\frac{1}{4} \left[ \log \frac{1+\sqrt{1-z}}{1-\sqrt{1-z}} - i \pi
\right ]^2 \;, & z <1.
\end{array}
\right .
\end{displaymath}

\end{document}